# Compound event metrics detect and explain ten-fold increase of extreme heat over Europe

**Gottfried Kirchengast[1,2,✉], Stephanie J. Haas[1] & Jürgen Fuchsberger[1]**

[1] Wegener Center for Climate and Global Change, University of Graz, Graz, Austria
[2] Institute of Physics, University of Graz, Graz, Austria
✉ corresponding author e-mail: gottfried.kirchengast@uni-graz.at

*Abstract:* Weather and climate extremes such as heatwaves are crucial climate hazards to people and ecosystems worldwide. In any region, climate change may alter their characteristics in complex ways so that rigorous and holistic quantification of the extremity of such events remains a challenge, impeding also uses by impact, attribution, and litigation communities. Here we introduce a new class of threshold-exceedance-amount metrics that track changes in event frequency, duration, magnitude, area, and timing aspects like daily exposure and seasonal shift—separately and up to compound total extremity (TEX). Applying them to extreme heat and showing their utility, at local- to country-scale (example Austria) and across European land regions, we revealed TEX amplifications of ~10 [4 to 25] (Europe 45-55°N) and ~8 [6 to 20] (Austria) for 2008-2022 vs. 1961-1990, strongly emerged from natural variability and an unequivocal evidence of anthropogenic climate change. Given their fundamental capacity to reliably track any threshold-defined climate hazard at any location, the new metrics enable a myriad of uses, from improving impact quantifications to climate action.

*Contents:*

***Main text***

## Introduction

The increase of weather and climate extremes, coming along with anthropogenic global warming and the associated gradual yet systematic climatic changes in all regions of the world, is known to be one of the inevitable consequences of climate change[1–3]. The related risks to societies and ecosystems are one of five major reasons for concern assessed by the IPCC[4,5]. In which ways climate change may affect the regional characteristics of these climate hazards such as heat waves, heavy precipitation events, droughts and wildfire events was succinctly summarized in the recent IPCC reporting on extreme events[2] and on informing impacts and risk assessment[3].

The main characteristics include increases in event frequency and duration as well as in exceedance magnitude and area (spatial extent), quantified in relation to a hazard threshold in a climate key variable[3] (e.g., surface air temperature) (Fig. 1a). In addition, characteristics of event timing, such as seasonal shifts in occurrence and related changes in annual exposure period or increases in daily exposure time, can be further traits exacerbating risks and damage potential[5–7] (e.g., earlier spring heat that leads to earlier fruit blossoms being more vulnerable to late frost events increasing yield risks in agriculture[8], or prolonged times of heat exposure worsening risks to human health and even survival[7,9,10], most strongly for vulnerable people who lack adaptive capacity[5,11]).

The IPCC assessments[2,3,12] and many further recent studies (e.g., ref. 11,13–20) have defined and used a wide range of extreme event indices, such as annual maximum value of daily maximum temperature $T^{\max}$ (TXx) or number of warm days above 90$^{\text{th}}$ percentile 1961-1990 (TX90p) for heat extremes[15]. These were explored separately and collectively and led to comprehensive and convincing evidence on recent past and projected future intensification especially of heat, precipitation, and drought-related extremes under climate change in many land regions of the world[1,2] and their mostly adverse impacts on ecosystems and people[3,5,21].

Regarding heat extremes, it was assessed for global land regions that the frequency of 10-year events (once-in-a-decade $T^{\max}$) increased about three-fold from 1850-1900 to the 2010s[1], while their intensity ($T^{\max}$ magnitude) increased by about 1.3 °C; for Europe an about two-fold increase in TX90p and an increase by about 2 °C in TXx was found from 1961-1990 to the 2010s[15].

Despite these advances and the role of heat-related extremes as prime climate hazards dominating global climate damage in measures like social cost of carbon[22] and human cost of global warming[23], a holistic methodology that jointly quantifies all key characteristics of extreme events in any region of interest remained a gap so far. It should unify fine-grained insight on individual characteristics with tracking of the compound extremity from the change of multiple characteristics, along with providing capability to quantify anthropogenic amplification versus natural variability.

This gap that we address in this study also impedes promising downstream uses of such more rigorous hazards quantification, for example, for better estimating anthropogenically caused shares in economic damage from weather extremes[5,22,24] or in human health impairment of vulnerable people from excess heat exposure[5,11,21], a capacity critically relevant for aims like climate change litigation towards compensation for damage or violation of human rights[7,25–27].

## New holistic threshold exceedance metrics for tracking extremes

Here we help close this gap and introduce and apply threshold-exceedance-amount (TEA) indicators based on long-term gridded key variable data as a new class of extremes-tracking metrics (for details see Methods, for illustration Fig. 1) that:
(1) consistently quantify changes in event frequency (EF), duration (ED), magnitude (EM), area (EA), daily exposure time (DET) and annual exposure period (AEP)—separately, partially compound as temporal events extremity (tEX) and event severity (ES), and as total events extremity (TEX); (2) flexibly focus on any georegion (GR) and annual climatic time period of interest (e.g., full year or warm season only); (3) typically use a grid-resolved high-percentile-based reference threshold map in a key climate variable to delineate grid values considered extreme; (4) also include amplification factors over the recent decades vs. a suitable reference period (e.g., 1961-1990) for all metrics, and optionally associated natural variability estimates; (5) and include aggregate-georegion (AGR) metrics, derived as means along with spread estimates from averaging over the individual GRs within an AGR (e.g., a continent-scale European land area).

We apply the new metrics (Fig. 1a-b), using daily maximum and hourly temperatures as key variables, to explore changes in heat extremes over Europe (Fig. 1c) under the recent climate change at a new holistic level of combined fine-grained (EF, ED, EM, EA; DET, AEP) and compound insight (tEX, ES, TEX). As a complement and to further demonstrate the utility of the metrics, also heavy precipitation extremes are inspected in an example subregion based on daily precipitation amount. The extremity amplification is quantified relative to the baseline extremity of the 1961-1990 reference period (main amplification factors $A^{\mathrm{F}}$, $A^{\mathrm{D}}$, $A^{\mathrm{M}}$, $A^{\mathrm{A}}$, $A^{\mathbf{t}}$, $A^{\mathbf{S}}$, $A^{\mathbf{T}}$; see Fig. 1b).

We first inspect cascaded local- to country-scale GRs for the example country Austria (Fig. 1d-e) based on key variable data at 1-10-km-scale resolution (SPARTACUS[28,29], ERA5-Land[30], ERA5[31], see Methods), comparing the degree of amplification revealed by the TEA metrics over the recent decades vs. 1961-1990 for this example country also to the associated range of natural variability (see Methods).

Subsequently we explore the heat extreme changes in regular-gridcell GRs (0.5° × 0.5° grid, with cell area around each grid point ~50,000 km$^2$) covering the entire European land area at 30-km-scale data resolution (ERA5[31], see Methods), from which we also derive the extreme heat amplification over European large-region AGRs and intercompare to the global-scale amplification of atmospheric heat gain due to global warming that was recently found near six-fold for the northern hemisphere extratropics for past-2000 vs. 1961-2000[32]. In the spirit of a compact study breaking ground for the new methodology, we chose this focus on Europe and extreme heat here, leaving further uses to follow-on studies (see Discussion).

The reference threshold maps for the GRs, which define where extreme values start based on a "reference climate"[2,3], were derived from the 1961-1990 data period as 99$^{\mathrm{th}}$ percentile of daily maximum temperature $T^{\mathrm{Max}}$ for delineating heat extremes (Fig. 1d-e, Extended Data Fig. 1a-c), and as 95$^{\mathrm{th}}$ percentile of the daily precipitation amount of warm-season wet days for precipitation extremes (Extended Data Fig. 1d) (see also Methods). In the GR Austria (Fig. 1d), and in particular in its southeastern Austria sub-GRs (Fig. 1d-e), the $T^{\mathrm{Max}}$ threshold appears overall similar to a constant hot-days threshold of 30 °C, while at the same time reliably capturing regional extreme-value variations (of mostly within 25 °C to 33 °C), primarily from surface altitude changes due to the distinct Austrian topography. Across Europe (Extended Data Fig. 1a), the threshold map as well reflects topography, but most saliently the north-to-south latitudinal $T^{\mathrm{Max}}$ gradient, leading to threshold-value variation from below 20 °C in some Scandinavian regions to above 35 °C in southern-Europe regions like in Spain or Greece. In general, any other suitable threshold formulation could be used as well for tailoring the TEAs to an application of interest (see Methods).

## Local- to country-scale amplification of extremes—example Austria

We find a near eight-fold amplification of the decadal-mean total events extremity ($A^{\mathbf{T}}$ ~7.9) of extreme heat events during the "Current climate" CC2008-2022 vs. the "Reference climate" Ref1961-1990 for the whole-country GR Austria, to which the frequency amplification contributed $A^{\mathrm{F}}$ ~1.7 while the severity amplification dominated with $A^{\mathbf{S}}$ ~4.7. (Fig. 2a,h). The largest individual factor in $A^{\mathbf{S}}$ is exceedance area expansion ($A^{\mathrm{A}}$ ~1.8) whereas event duration and exceedance magnitude amplified similarly to the event frequency by ~1.6-1.7 (Fig. 2a-d). The local-scale tracking, mapped at the 1-km-scale input data grid for the temporal-only metrics $A^{\mathrm{F}}$, $A^{\mathrm{D}}$, $A^{\mathrm{M}}$ (Fig. 2e-g), reveals significant spread for local extreme heat amplification within the GR that reaches well beyond the average GR values up to factors of ~3-5, indicating the value of this fine-grained mapping for downstream uses at the heat extremes impact scale of local settlements of people, natural habitats and ecosystems.

Comparing this degree of decadal-mean amplification to the upper-half 90% CI range of natural variability ($1+1.645{\cdot}s^{\mathrm{upp}}$), estimated from centennial station datasets in the GR[33] (Extended Data Fig. 2) and backed by evidence from a recent 1421-2008 multicentury reanalysis[34] (see Methods), we notice that the individual frequency, duration, and area amplifications have emerged from the range (probability $p_{\mathrm{natural}} < 5\,\%$) only past 2000 (magnitude mid-1990s), to roughly a 3σ-level ($1+3{\cdot}s^{\mathrm{upp}}$) in CC2008-2022 (Fig. 2a-d). However, the TEX amplification $A^{\mathbf{T}}$ (Fig. 2h) emerged in the early 1990s already and has in CC2008-2022 reached well beyond an estimated 10σ-level ($p_{\mathrm{natural}} \cong 0$), a change impossible due to natural variability. It is hence unequivocally clear that this strong increase in extreme heat over Austria was caused by the recent climate change primarily driven by anthropogenic greenhouse gas and aerosol net-forcing increase[1,32,35,36].

Inspecting both the native annual and decadal-mean annual metrics directly (see Extended Data Fig. 3), in addition reveals: (1) their changes since the 1960s in actual physical units, from event no. / yr (EF) to total areal °C days / yr (TEX); (2) the strong year-to-year variations of heat extremes, typical for continental Europe and many other land regions worldwide[2,15,16], indicating the advantage of rolling decadal-means as main indicators while still co-tracking the spread of the individual years around the mean; (3) the differentiated character of individual extreme heat years, like the three record extreme years as gauged by the TEX—2003, 2013, 2015—showing significantly different mixes of event frequency and severity characteristics; (4) the basic capacity of the new metrics to quantify also the extremity of individual years (and even of individual strong events), in terms of the degree of exceeding the range of natural variability and hence of the estimated share attributable to anthropogenic climate change.

Inspecting as well heavy precipitation in the GR SEA subregion in the same way as the heat extremes (Fig. 2), as a complementary major type of extreme with very different character involving much more local-scale variability[37,38], we find (Extended Data Fig. 4): (1) the basic event frequency, duration, and magnitude amplifications much smaller ($A$s ~1.07-1.15) and in CC2008-2022 still within (or very close to) the estimated range of natural variability; (2) the compound tEX amplification ($A^{\mathbf{t}}$ ~1.35) as a signal appearing close to emergence from natural variability during the 2010s while still being confounded with substantial local-scale and natural variability[29,39].

We next explore the heat extremes in the GR AUT and its sub-country GRs SEA and FBR for closer insight based on multiple-gridscale data (SPARTACUS, ERA5-Land, ERA5), again including heavy precipitation as further extreme in the sub-country GRs, and additionally inspect the supplementary metrics daily exposure time (DET) and annual exposure period (AEP) for heat extremes. Based on cascaded (dis)aggregation of the TEA amplification metrics, we find the heat amplification results (Fig. 3a-c; Extended Data Fig. 5a-c) reasonably consistent among the three different datasets in the three GRs, with the amplifications becoming stronger in the smaller southeastern Alpine foreland GRs (e.g., $A^{\mathbf{T}}$ ~16-20 for FBR vs. $A^{\mathbf{T}}$ ~6-8 for AUT)—in line with this region being sensitive to climate-change-induced summer warming[35,40–42].

The precipitation amplification results (Fig. 3d-f; Extended Data Fig. 5d-f) reveal, while ERA5-Land shows some skill to grossly reflect the tEX amplification pattern seen by SPARTACUS across SEA (Fig. 3d), that the 10-km-scale datasets are severely limited in capturing the strong local-scale variability in these small GRs—in line with the findings in a recent detailed study[39]. None of the datasets can reliably capture a change signal beyond natural variability, apart from how the tEX amplification appears close to emergence in the GR SEA (Fig. 3e) as discussed before (Extended Data Fig. 4d).

The daily exposure time to extreme heat (h/day) and its amplification $A^{\mathrm{h}}$ (Extended Data Fig. 6a,b), explored based on the hourly ERA5 data, show a clear increase in all three GRs, again with stronger increase in the smaller GRs ($A^{\mathrm{h}}$ ~1.25-1.75 in CC2008-2022). Factoring also this sub-daily metric of heat intensification into the main TEX (yielding $A^{\mathbf{hT}} = A^{\mathrm{h}} \cdot A^{\mathbf{T}}$) is, for example, vital for accurate assessment of health impacts on heat-vulnerable persons[5,7,11] or populations[7,9,10]. Its inclusion further increases the estimated SPARTACUS-based $A^{\mathbf{T}}$ for the GRs (Fig. 3a-c; $A^{\mathbf{T}}$ ~8-18) to $A^{\mathbf{hT}}$ = ~10-31, a striking ten- to thirty-fold increase of extreme heat exposure in these central-Europe GRs.

The annual exposure period and its amplification $A^{\Delta\mathrm{Y}}$, inspected based on ERA5 and SPARTACUS (Extended Data Fig. 6c-f), indicate a clear expansion of the annual extreme heat season in all three GRs, with an increase by about a month since 1961-1990 (due to both earlier first and later last events per year), corresponding to estimated amplifications of ~1.35-1.75 (AUT) to ~2.3-2.75 (FBR).

## Amplification of heat extremes and atmospheric heat over Europe

Understanding the utility of the new metrics for consistent fine-grained and compound insight from the example GR Austria, we now apply them based on ERA5 to the gridded GRs of similar size across Europe (Fig. 1c). We first inspect the amplification map of heat extremes in compound TEX ($A^{\mathbf{T}}$; Fig. 4a) and its components ($A^{\mathrm{F}}$, $A^{\mathrm{D}}$, $A^{\mathrm{M}}$, $A^{\mathrm{A}}$; Extended Data Fig. 7a-d) for CC2008-2022 vs. Ref1961-1990.

We find strongest $A^{\mathbf{T}}$s ~10 in continental Europe (with hotspot-GRs > 20 in Ukraine and surrounding east-Europe regions), $A^{\mathbf{T}}$s ~6 in southern Europe (with hotspot-GRs > 12 in southeast-Europe Balkan countries and Spain), and clearly smaller $A^{\mathbf{T}}$s ~2.5 in northern Europe (with GRs < 1 in northern Scandinavia and central Sweden). Frequency amplification is generally the strongest component (Extended Data Fig. 7a), while the

hotspot-$A^{\mathbf{T}}$s are further enhanced especially by elevated duration and magnitude amplifications in these regions (Extended Data Fig. 7b,c).

Taking next a fine-grained look at time-dynamic TEX amplification trajectories $A^{\mathbf{T}}{}_{s}(t)$ over the core years 1966 to 2018 in amplification-factor charts of Fig.1b-type for three representative GRs across Europe (Fig. 4b), we see a strong emergence of the TEX amplification signal, especially in the SAF and IBE GRs. The accelerated pace over the recent two decades from 2000 (reddish core years 2006 to 2018) is particularly striking, with the CC2008-2022 signal (dark-red dot with associated light-reddish area) strongly above natural variability (proxy-indicated by the "cloud" of gray core years 1966 to 1990), most so in the central-European GR SAF, consistent with the results for the SAF-related Austrian GRs (Fig. 3a-c).

We close this inspection with an aggregate view on the TEX amplification over the continent-scale AGRs of the entire European land and its central, southern, and northern parts (EUR, C-EUR, S-EUR, N-EUR; see Fig. 4a), for AGR-mean estimates and the associated 90% CI spread of GR amplifications around the mean (Fig. 4c). For C-EUR, which represents continental-land conditions, we find an $A^{\mathbf{T}}$ ~10 [4 to 25] in CC2008-2022 vs. Ref1961-1990. The mean for EUR overall (~6) lies close to the S-EUR mean and the one for N-EUR is distinctly lower, consistent with the map in Fig. 4a. The ten-fold C-EUR mean amplification is about a factor of five stronger than the amplification vs. Ref1961-1990 over Europe found for the more limited warm-days metric TX90p[15] (corresponding to $A^{F} \cdot A^{D}$, and 90$^{th}$ instead of 99$^{th}$ percentile). This underpins the added value of compound metrics such as TEX as part of the new class of metrics.

We finally use the TEX to check how the near-surface extreme heat amplification, in terms of the increase in the aggregate threshold exceedance energy deposited annually in the atmospheric boundary layer (ABL), scales to the bulk atmospheric heat increase from global warming. The gain in atmospheric heat content (AHCg) due to the global warming was recently found to have amplified near six-fold in the northern hemisphere extratropics for 2001-2020 (~64 EJ/yr) vs. 1961-2000 (~11 EJ/yr)[32], with ERA5 as one of the reanalyses involved having shown a closely consistent result. One may expect, due to extreme warming basically changing stronger than bulk warming[1,2], that the near-surface exceedance energy gain from heat extremes has amplified even stronger than AHCg (here computed from ERA5, as the TEX) over the land regions of Europe.

We use the annual ABL exceedance heat content (AEHC)—proportional to the TEX via the approximative ABL excess heat uptake capacity (see Methods)—as reasonable proxy for this exceedance energy gain (in PJ/yr) and indeed find the AEHC amplification over the continental land as represented by C-EUR distinctly stronger than the one of AHCg (Fig. 4d). Specifically, the C-EUR AEHC exhibits an average eight- to ten-fold amplification in 2001-2020|2008-2022 (~1150|1380 PJ/yr) vs. 1961-1990 (~140 PJ/yr), about twice as much as the NH35-70N warm-season AHCg that was amplified about four- to six-fold during this time (consistent with ref. 32; the 1961-2000 baseline led to somewhat larger amplifications therein). Enabling climate change physics checks like this further underpins the value of the TEX metric.

## Discussion and perspectives

In this Article we have introduced a new holistic class of threshold-exceedance-amount (TEA) indicators that can track the multiple characteristics of extreme events under climate change, and of any other threshold-defined hazards, with a level of rigor and consistency in both fine-grained and compound insight not attained before. Using it to explore heat extremes over Europe from the 1960s to the present in a cascade of local- to continental-scale regions, also peering into precipitation extremes in the smaller-scale regions, showed its profound capacity beyond the existing more dispersed variety of extreme event indices[2,12–15] and their side-by-side rather than holistic use in the recent IPCC reporting[1–3].

Along with fine-grained explanatory insight into changes in event frequency, duration, magnitude, areal extent, and daily exposure time we found, based on 99$^{th}$-percentile daily-max-temperatures 1961-1990 as reference threshold, a more than eight-fold amplification of the decadal-mean total events extremity (TEX) for 40 % of the European land area, including more than ten-fold for 57 % of the C-EUR area. This severe heat increase under the recent climate change appears strongly emerged from natural variability and about twice as strong as the overall northern mid-latitude atmospheric heat increase due to global warming. The associated increase in risk and harmful impacts is strongly evidenced by recent assessments[3,5,7] and many specific studies such as on agricultural loss[22,43] and heat mortality[9–11,22].

Based on this and in perspective, the climate science community and practice users in the climate services and action domains (e.g., hydro-meteorological services, environmental agencies, insurance companies, law firms, public administrations, climate policymakers) can build on our results and deploy the versatility and diagnostic power of the TEA metrics in a much wider array of applications, such as:

(1) tracking and comprehensively quantifying heat and precipitation extremes and further key extremes like droughts, wildfires, flooding, and storminess in all georegions of interest worldwide, over the recent past (e.g., using ERA5, ERA5-Land or similar reanalyses and region-specific datasets) and the future (e.g., using weather forecasts and seasonal predictions[44,45], and CMIP6[46] and CORDEX[47] climate model projections);

(2) computing TEA metrics in tailored form for serving risk and impact analyses for any climate hazard of concern in georegions of interest and integrating them with exposure and vulnerability geodata, to obtain time-dynamic risk and impacts tracking and forecasting for threats like socio-economic loss (e.g., estimating damage costs from drought[48] or flooding[24]) and harm to people (e.g., estimating deadly heat risk to whole regions[9,10] or health impairment risk to vulnerable persons[7,11,21]);

(3) extreme events attribution[49–51], based on computing the climate hazard indicators of interest (e.g., the TEX of a georegion or the tEX of a location) in suitable form along with their associated natural variability, to quantify the share of the hazard extremity (of a decade, year, or single exceptional event) that is attributable to anthropogenic climate change, optionally also the corresponding share in the estimated damage to properties or harm to people (e.g., for possibly claiming compensation through litigation[25,26,52]);

(4) any further applications beyond weather and climate extremes (e.g., for exploring TEAs also in socio-economic or ecosystem change processes), since the metrics class is derived through a generically applicable space-time filter for cascaded and multi-scale extraction of extremes and their statistics along time in threshold-bound subspaces of a spatiotemporal field of any metric-scaled key variable of interest (see Methods).

As with all types of data-driven climate change metrics, there are also inevitable limitations to the practical utility of the TEA indicators. Most importantly, the reliability of the results rests on sufficient quality (spatiotemporal resolution, adequate accuracy, long-term stability) of the gridded-field input data of the key variable used to extract the metrics. Climate-quality datasets are hence required, in particular if high-percentile thresholds are employed, including for datasets optionally used to co-estimate natural variability.

Furthermore, the sensitivity of results to the choice of reference threshold map should be well understood for any use, since this map fundamentally delineates what is "extreme", the desired TEA subspace, from what is "regular" and discarded; for very rare extremes (like > p99.8) extreme-value-theory (EVT) methods[14,53,54] may be preferable. Natural variability estimation should follow proven methods[49,51], especially if serving climate change attribution for downstream uses[52,55].

While acknowledging cautions like these we at the same time witness a steady increase in climate-quality datasets both from observations and models[46,47,56,57]—from local- to global-scale and for all parts of the climate system. We hence suggest broadest possible use of the added value of the new metrics class, for helping better cope with climate hazards and their risks and impacts, and for promoting climate action in line with the Paris Agreement to prevent further acceleration of the global threats they pose.

# References

1. IPCC 2021 Summary for Policymakers. in *Climate Change 2021: The Physical Science Basis* (eds Masson-Delmotte, V. et al.) 3–32 (Cambridge Univ. Press, 2023).
2. Seneviratne, S. I. et al. Weather and climate extreme events in a changing climate. in *Climate Change 2021: The Physical Science Basis* (eds Masson-Delmotte, V. et al.) 1513–1766 (Cambridge Univ. Press, 2023).
3. Ranasinghe, R. et al. Climate change information for regional impact and for risk assessment. in *Climate Change 2021: The Physical Science Basis* (eds Masson-Delmotte, V. et al.) 1767–1926 (Cambridge Univ. Press, 2023).
4. O'Neill, B. et al. IPCC reasons for concern regarding climate change risks. *Nat. Clim. Change* **7**, 28–37 (2017).
5. IPCC 2022 Summary for Policymakers. in *Climate Change 2022: Impacts, Adaptation, and Vulnerability* (eds Pörtner, H.-O. et al.) 3–33 (Cambridge Univ. Press, 2023).
6. Suarez-Gutierrez, L., Müller, W. A. & Marotzke, J. Extreme heat and drought typical of an end-of-century climate could occur over Europe soon and repeatedly. *Commun. Earth Environ.* **4**, 415 (2023).
7. van Daalen, K. R. et al. The 2024 Europe report of the *Lancet* Countdown on health and climate change: unprecedented warming demands unprecedented action. *Lancet Publ. Health* **9**, e495–e522 (2024).
8. Unterberger, C. et al. Spring frost risk for regional apple production under a warmer climate. *PLoS ONE* **13**, e0200201 (2018).
9. Mora, C. et al. Global risk of deadly heat. *Nat. Clim. Change* **7**, 501–506 (2017).
10. Vicedo-Cabrera, A. M. et al. The burden of heat-related mortality attributable to recent human-induced climate change. *Nat. Clim. Change* **11**, 492–500 (2021).
11. Brimicombe, C. et al. Preventing heat-related deaths: The urgent need for a global early warning system for heat. *PLOS Clim.* **3**, e0000437 (2024).
12. Gutiérrez, J. M. et al. Annex VI: Climatic impact-driver and extreme indices. in *Climate Change 2021: The Physical Science Basis* (eds Masson-Delmotte, V. et al.) 2205–2214 (Cambridge Univ. Press, 2023).
13. Zhang, X. et al. Indices for monitoring changes in extremes based on daily temperature and precipitation data. *WIREs Clim. Change* **2**, 851–870 (2011).
14. Schär, C. et al. Percentile indices for assessing changes in heavy precipitation events. *Climat. Change* **137**, 201–216 (2016).
15. Dunn, R. J. H. et al. Development of an updated global land in situ-based data set of temperature and precipitation extremes: HadEX3. *J. Geophys. Res. Atmos.* **125**, e2019JD032263 (2020).
16. Perkins-Kirkpatrick, S.E. & Lewis, S.C. Increasing trends in regional heatwaves. *Nat. Commun.* **11**, 3357 (2020).
17. Di Napoli, C., Barnard, C., Prudhomme, C., Cloke, H. L. & Pappenberger, F. ERA5-HEAT: A global gridded historical dataset of human thermal comfort indices from climate reanalysis. *Geosci. Data J.* **8**, 2–10 (2021).
18. Song, F., Zhang, G. J., Ramanathan, V. & Leung, L. R. Trends in surface equivalent potential temperature: A more comprehensive metric for global warming and weather extremes. *Proc. Natl Acad. Sci. USA* **119**, e2117832119 (2022).
19. Brimicombe, C. et al. Wet Bulb Globe Temperature: Indicating extreme heat risk on a global grid. *GeoHealth* **7**, e2022GH000701 (2023).
20. Li, Y., Kirchengast, G., Schwaerz, M. & Yuan, Y. Monitoring sudden stratospheric warmings under climate change since 1980 based on reanalysis data verified by radio occultation. *Atmos. Chem. Phys.* **23**, 1259–1284 (2023).
21. Thiery, W. et al. Intergenerational inequities in exposure to climate extremes. *Science* **374**, 158–160 (2021).

22. Rennert, K. et al. Comprehensive evidence implies a higher social cost of $CO_2$. *Nature* **610**, 687–692 (2022).
23. Lenton, T.M. et al. Quantifying the human cost of global warming. *Nat. Sustain.* **6**, 1237–1247 (2023).
24. Unterberger, C., Hudson, P., Botzen, W. J. W., Schroeer, K. & Steininger, K. Future public sector flood risk and risk sharing arrangements: An assessment for Austria. *Ecol. Econ.* **156**, 153–163 (2019).
25. Schulev-Steindl, E., Hinteregger M., Kirchengast, G. et al. (Eds) *Climate Change, Responsibility and Liability*. 525 pp (Nomos, 2022).
26. Phelan, A. L. et al. Collective action and legal mobilisation for the right to health in the climate crisis. *Lancet* **403**, 2272-2274 (2024).
27. Blattner, C. E. European ruling on climate and rights is a game changer. *Nature* **628**, 691 (2024).
28. Hiebl, J. & Frei, C. Daily temperature grids for Austria since 1961—concept, creation and applicability. *Theor. Appl. Climatol.* **124**, 161–178 (2016).
29. Hiebl, J. & Frei, C. Daily precipitation grids for Austria since 1961—development and evaluation of a spatial dataset for hydroclimatic monitoring and modelling. *Theor. Appl. Climatol.* **132**, 327–345 (2018).
30. Muñoz-Sabater, J. et al. ERA5-Land: a state-of-the-art global reanalysis dataset for land applications. *Earth Syst. Sci. Data* **13**, 4349–4383 (2021).
31. Hersbach, H. et al. The ERA5 global reanalysis. *Q. J. Roy. Meteor. Soc.* **146**, 1999–2049 (2020).
32. von Schuckmann, K., Minière, A., Gues, F., Cuesta-Valero, F. J., Kirchengast, G. et al. Heat stored in the Earth system 1960–2020: where does the energy go?, *Earth Syst. Sci. Data* **15**, 1675–1709 (2023).
33. GeoSphere Austria – Measurement stations Daily data v2 datasets. Online at GeoSphere Data Hub https://doi.org/10.60669/gs6w-jd70 (2024).
34. Valler, V. et al. ModE-RA: a global monthly paleo-reanalysis of the modern era 1421 to 2008. *Sci. Data* **11**, 36 (2024).
35. Schumacher, D. L. et al. Exacerbated summer European warming not captured by climate models neglecting long-term aerosol changes. *Commun. Earth Environ.* **5**, 182 (2024).
36. Rousi, E., Kornhuber, K., Beobide-Arsuaga, G., Luo, F. & Coumou, D. Accelerated western European heatwave trends linked to more-persistent double jets over Eurasia. *Nat. Commun.* **13**, 3851 (2022).
37. Schroeer, K. & Kirchengast, G. Sensitivity of extreme precipitation to temperature: the variability of scaling factors from a regional to local perspective. *Clim. Dyn.* **50**, 3981–3994 (2018).
38. Schroeer, K., Kirchengast, G. & O, S. Strong dependence of extreme convective precipitation intensities on gauge network density. *Geophys. Res. Lett.* **45**, 8253–8263 (2018).
39. Haas, S. J., Kirchengast, G. & Fuchsberger, J. Exploring possible climate change amplification of warm-season precipitation extremes in the southeastern Alpine forelands at regional to local scales. *J. Hydrol. Reg. Studies* **56**, 101987 (2024).
40. Auer, I. et al. HISTALP—historical instrumental climatological surface time series of the Greater Alpine Region. *Int. J. Climatol.* **27**, 17–46 (2007).
41. Kabas, T., Foelsche, U. & Kirchengast, G. Seasonal and annual trends of temperature and precipitation within 1951/1971-2007 in south-eastern Styria, Austria. *Meteorol. Z.* **20**, 277–289 (2011).
42. Hohmann, C., Kirchengast, G. & Birk, S. Alpine foreland running drier? Sensitivity of a drought vulnerable catchment to changes in climate, land use, and water management. *Climat. Change* **147**, 179–193 (2018).
43. Zhao, C. et al. Temperature increase reduces global yields of major crops in four independent estimates. *Proc. Natl Acad. Sci. USA* **114**, 9326–9331 (2017).
44. Brown, A. et al. Unified modeling and prediction of weather and climate: A 25-year journey, *Bull. Amer. Meteor. Soc.* **93**, 1865–1877 (2012).

45. Johnson, S. J. et al. SEAS5: the new ECMWF seasonal forecast system. *Geosci. Model Dev.* **12**, 1087–1117 (2019).
46. Eyring, V. et al. Overview of the Coupled Model Intercomparison Project Phase 6 (CMIP6) experimental design and organization. *Geosci. Model Dev.* **9**, 1937–1958 (2016).
47. Gutowski Jr., W. J. et al. WCRP COordinated Regional Downscaling EXperiment (CORDEX): a diagnostic MIP for CMIP6. *Geosci. Model Dev.* **9**, 4087–4095 (2016).
48. García-León, D., Standardi, G. & Staccione, A. An integrated approach for the estimation of agricultural drought costs. *Land Use Pol.* **100**, 104923 (2021).
49. van Oldenborgh, G.J. et al. Pathways and pitfalls in extreme event attribution. *Climat. Change* **166**, 13 (2021).
50. Gilford, D. M., Pershing, A., Strauss, B. H., Haustein, K. & Otto, F. E. L. A multi-method framework for global real-time climate attribution. *Adv. Stat. Clim. Meteorol. Oceanogr.* **8**, 135–154 (2022).
51. Otto, F. E. L. Attribution of extreme events to climate change. *Ann. Rev. Environ. and Res.* **48**, 813–828 (2023).
52. Stuart-Smith, R. F. et al. Filling the evidentiary gap in climate litigation. *Nat. Clim. Change* **11**, 651–655 (2021).
53. Naveau, P., Hannart, A. & Ribes, A. Statistical methods for extreme event attribution in climate science. *Ann. Rev. Stat. and Appl.* **7**, 89–110 (2020).
54. Thompson, V. et al. The most at-risk regions in the world for high-impact heatwaves. *Nat. Commun.* **14**, 2152 (2023).
55. Stott, P. A. & Christidis, N. Operational attribution of weather and climate extremes: what next? *Environ. Res. Clim.* **2**, 013001 (2023).
56. Dunn, R. J. H. et al. Global Climate – Appendix 2: Datasets and sources. in *State of the Climate in 2023* (eds Blunden, J. & Boyer, T.). *Bull. Amer. Meteor. Soc.* **105**, S124–S134 (2024).
57. Buontempo, C. et al. The Copernicus Climate Change Service: Climate science in action. *Bull. Amer. Meteor. Soc.* **103**, E2669–E2687 (2022).

## Figures

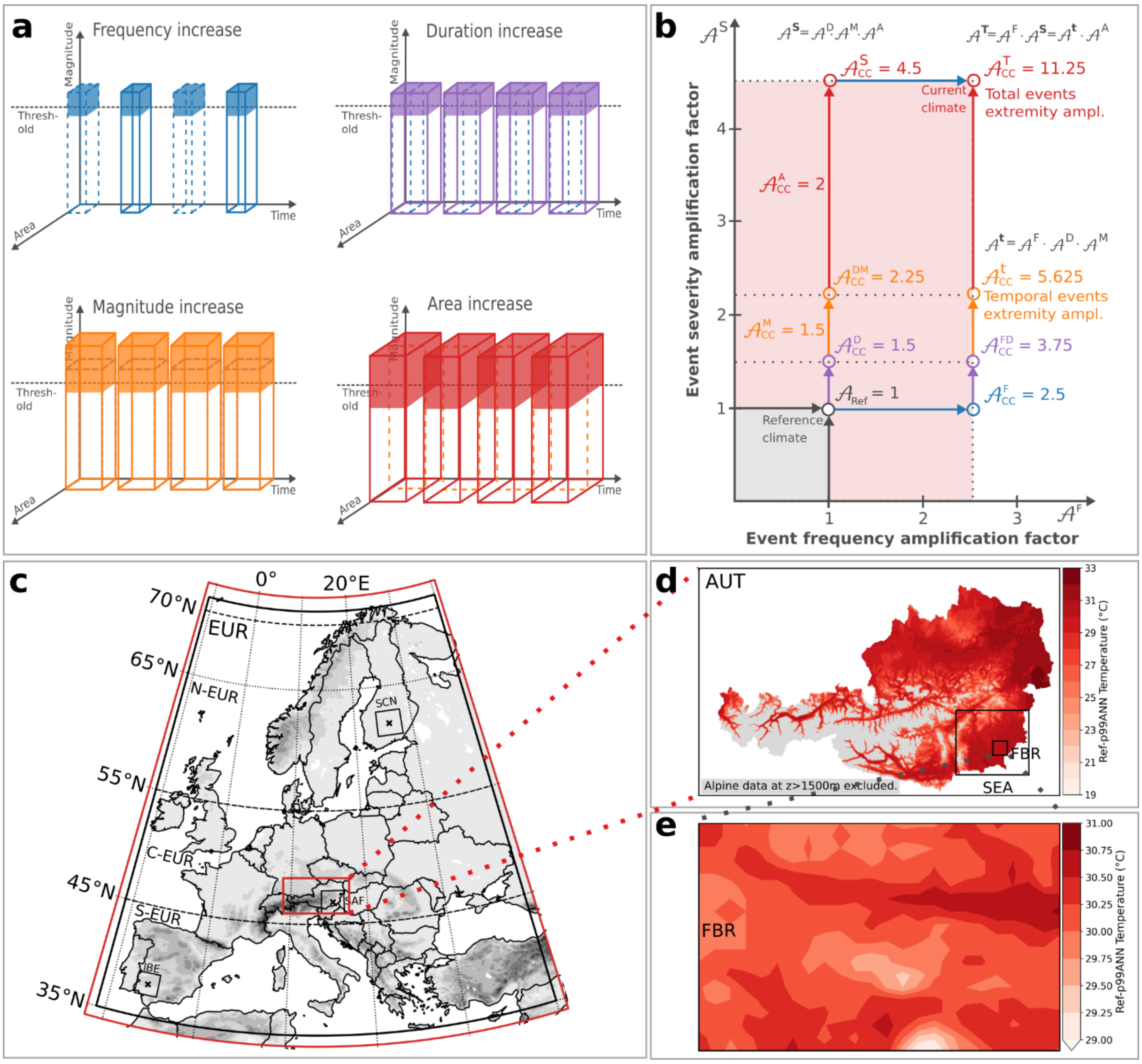

**Fig. 1 | New holistic class of metrics for tracking extreme event amplification under climate change.** **a**, Illustration of event frequency, event duration, exceedance magnitude and exceedance area increase under climate change, the four basic threshold-exceedance-amount (TEA) metrics of extreme events that factor into the TEA volume of total extremity (including all volumes above threshold in a “Current climate” vs. dashed-marked parts only in a “Reference climate”); further expansion of this TEA volume could come from additional extremity dimensions such as daily exposure. **b**, Normalizing to “Reference climate” metrics, an amplification-factors chart provides fine-grained and cascaded compound insight into amplification shares in a “Current climate”, for the dimensions of frequency amplification and severity amplification (including all per-event metrics; here duration × magnitude × area), with total events extremity amplification as the all-up result (example $A^{\mathrm{T}} = 11.25$). For any factor $A$ its lower-left area in the chart (here marked light-red for $A^{\mathrm{T}}$) corresponds to its value (i.e., the degree of amplification vs. the gray unit area $A_{\mathrm{Ref}} = 1$). **c-e**, The new metrics are applied to explore the amplification of heat extremes vs. Ref1961-1990 for gridded GRs (~50,000 km$^2$ size) across the entire European land area (**c**) and for Austrian example GRs embedded in central Europe (**d**) (whole country AUT, ~67,000 km$^2$ at altitudes ≤ 1500 m; Southeast Austria SEA ~9,600 km$^2$; Feldbach Region FBR, ~410 km$^2$ within SEA); 99$^{\mathrm{th}}$-percentile $T^{\mathrm{Max}}$ threshold maps are co-illustrated for AUT (**d**) and FBR (**e**) based on the 1-km-scale SPARTACUS data.

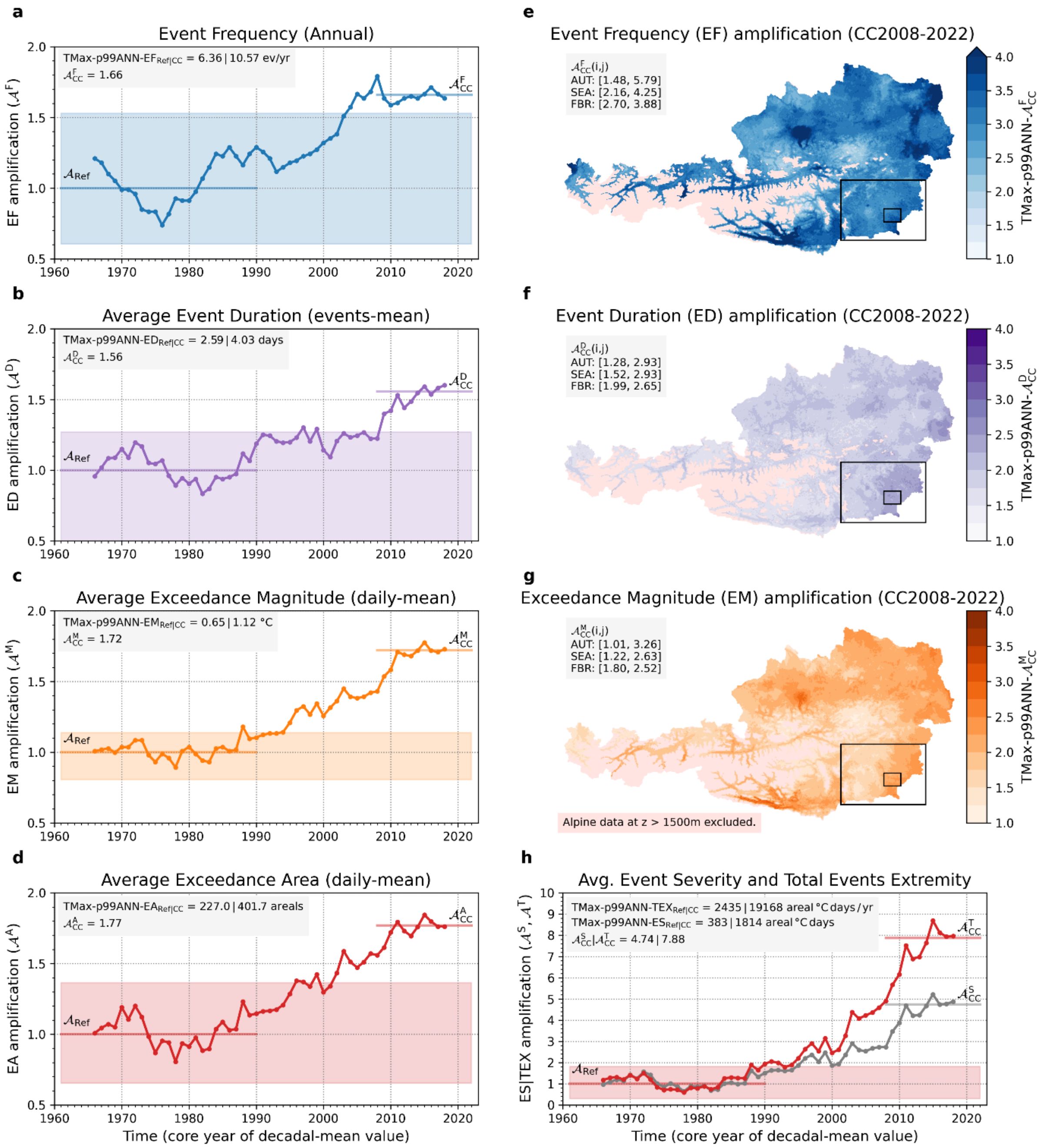

**Fig. 2 | Tracking the amplification of extreme heat events in the country GR Austria. a-d**, Decadal-mean annual amplification factor time series of the four basic TEA metrics frequency (**a**), duration (**b**), magnitude (**c**) and area (**d**) for heat extremes in the GR AUT based on SPARTACUS data and the threshold map shown in Fig. 1d (decade core years 1966 to 2018; reference period Ref1961-1990 with $\mathcal{A}_{Ref} = 1$; current climate period CC2008-2022, with $\mathcal{A}_{CC}$ value noted in panel legend). The estimated natural variability is shown as shaded corridor (90% CI) and the absolute values of the TEA metrics for the Ref and CC periods are noted in the panel legends (including one custom unit: 1 areal = 100 km$^2$). **e-g**, Maps of the local-scale amplification factors of the three basic temporal TEA metrics frequency (**e**), duration (**f**) and magnitude (**g**) in the CC2008-2022 period at the 1-km resolution of the SPARTACUS data. [Minimum, Maximum] values of all GR grid points are co-listed in panel legends for the GR AUT and its sub-GRs SEA and FBR (rectangular boxes). **h**, Amplification time series of the two compound TEA metrics event severity (ES = ED × EM × EA) and total events extremity (TEX = EF × ES) for the GR AUT (same format as **a-d**; gray curve $\mathcal{A}^S$, red curve $\mathcal{A}^T$; shaded corridor corresponding to $\mathcal{A}^T$).

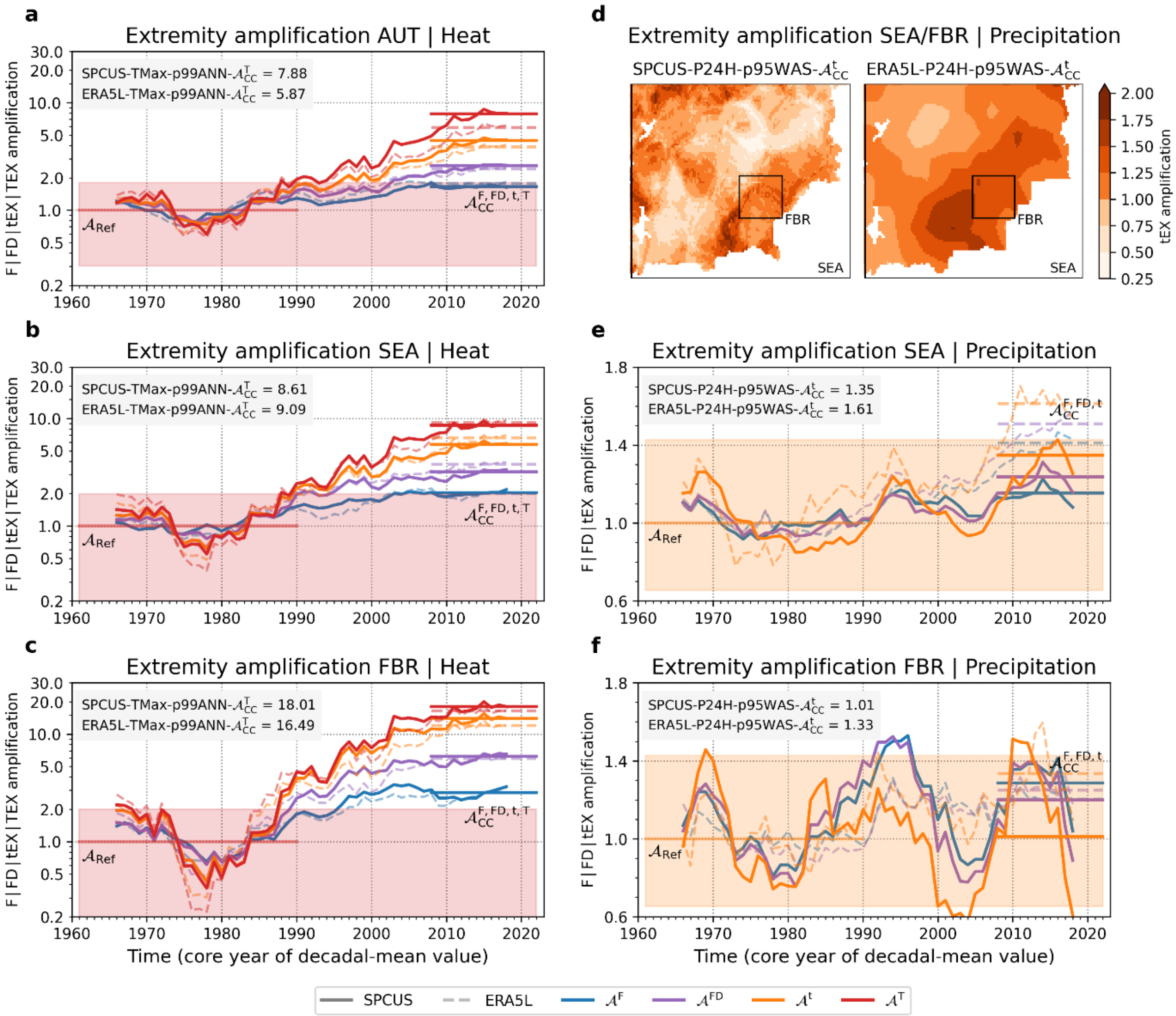

**Fig. 3 | Fine-grained insight to extreme heat and precipitation amplification based on multiple-gridscale data in the example GRs of Austria. a-c**, Decadal-mean annual amplification factor timeseries (as in Fig. 2a-d,h) of the cascade of (dis)aggregated TEA amplification metrics from frequency to total extremity ($A^F \rightarrow A^F{\cdot}A^D = A^{\mathbf{FD}} \rightarrow A^{\mathbf{FD}}{\cdot}A^M = A^t \rightarrow A^{\mathbf{t}}{\cdot}A^A = A^{\mathbf{T}}$; cf. Fig. 1b), for heat extremes in the GRs AUT (**a**), SEA (**b**) and FBR (**c**) based on 1-km-scale SPARTACUS (SPCUS, solid lines) and 10-km-scale ERA5-Land (ERA5L, dashed lines) data and the associated threshold maps shown in Fig. 1d and Extended Data Fig. 1b. The cascaded metrics are shown in log-space, due to the strong and non-linear amplification of the heat extremes (logarithmic axis, such as common in engineering for strong amplifier gains; see Methods). **d**, GR SEA (including FBR) local-scale maps of tEX amplification $A^t$ for precipitation extremes (left SPCUS, right ERA5L). **e**,**f**, Cascade of amplification metrics for precipitation extremes (temporal-only metrics up to $A^t$) in the GRs SEA (**e**) and FBR (**f**), shown in the same form as in **a-c** but here using a standard linear-scale axis, since the amplification is comparatively weak (within a factor of 2). The shaded corridors are the estimated natural variability ranges (90% CI) of the TEX amplification $A^T$ (red-shaded, **a-c**) or tEX amplification $A^t$ (orange-shaded, **e**,**f**), respectively.

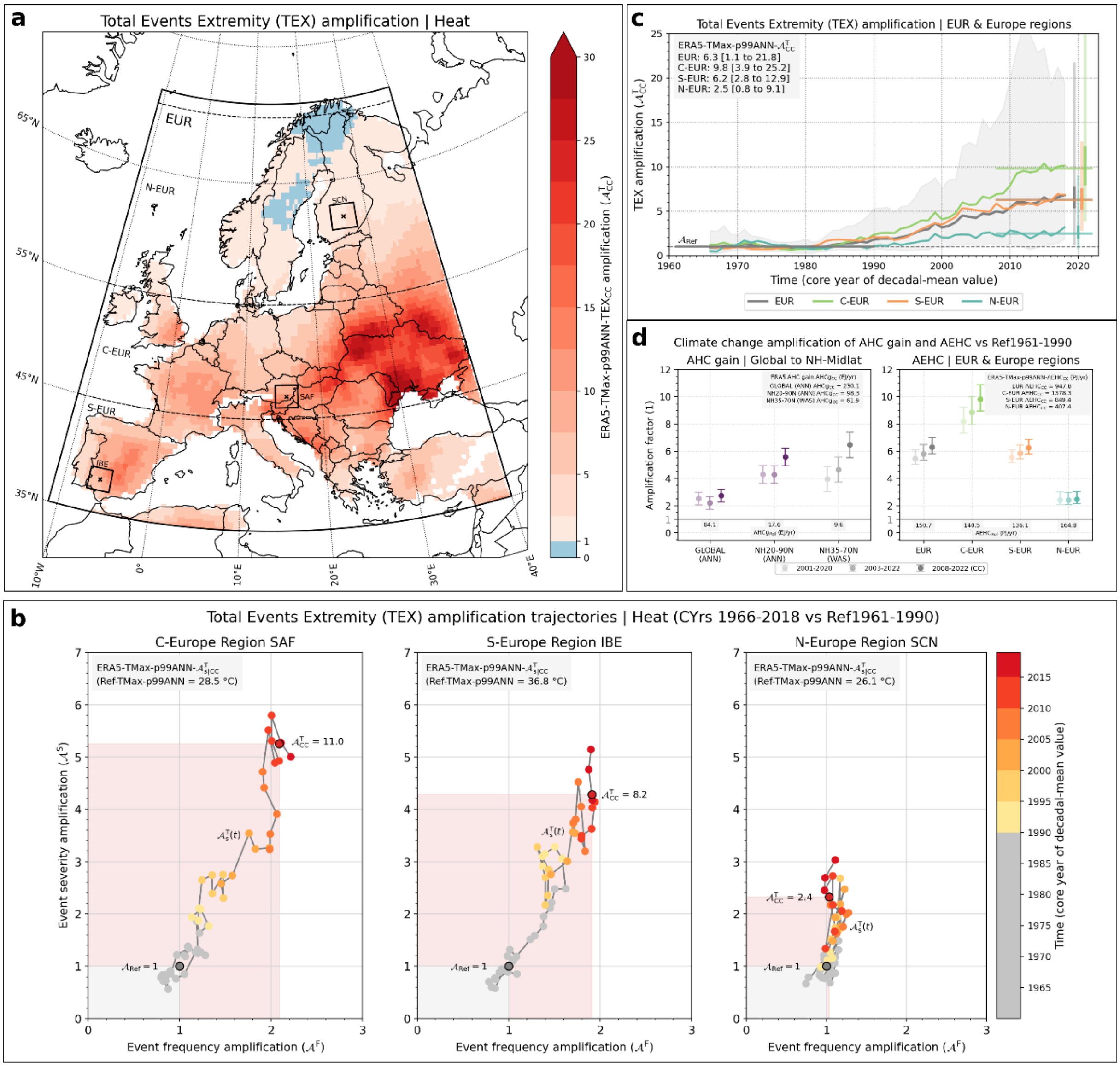

**Fig. 4 | Amplification of heat extremes and atmospheric heat over Europe.** **a**, Map of TEX amplification for CC2008-2022 vs. Ref1961-1990 in GRs across the European land area (EUR; 0.5° × 0.5° grid; GR area around grid points ±1° × ±1°/cos(latitude), ~50,000 km$^2$; land fraction ≤ 1500 m at least 0.5) based on ERA5 data and the threshold map in Extended Data Fig. 1a. Representative central Europe (C-EUR, 45-55 °N), southern Europe (S-EUR, 35-44.5 °N) and northern Europe (N-EUR, 55.5-70 °N) GRs are indicated as boxes with the GR center (grid point) marked as cross (Southeastern Alpine Forelands, SAF; Iberian, IBE; Scandinavian, SCN). **b**, Decadal-mean TEX amplification trajectories $A^{\mathrm{T}}{}_{s}(t)$ over the core years 1966 to 2018 in amplification-factor charts ($A^{\mathrm{F}} \times A^{\mathrm{S}}$, where $A^{\mathrm{S}} = A^{\mathrm{D}} \cdot A^{\mathrm{M}} \cdot A^{\mathrm{A}}$; see Fig. 1b) for the GRs SAF, IBE and SCN. The $A^{\mathrm{T}}{}_{\mathrm{CC}}$ value is shown as well (larger dark-red dot with related light-reddish area) and the GRs average threshold temperature is noted in the panel legends. **c**, Decadal-mean TEX amplification time series for the continental-scale AGRs EUR, C-EUR, S-EUR and N-EUR, indicating also (both as 90% CI) the spread of the individual GR values around the mean (shading and bars; values in legend) and the estimated uncertainty of the mean (darker color in bars). **d**, Comparison of the AEHC increase over the European AGRs, corresponding to the TEX amplification, to the overall global and mid-latitude AHC gain amplification, indicating also the values for the Ref and CC periods (legends) and estimated uncertainties (90% CI; bars).

## Extended Data Figures

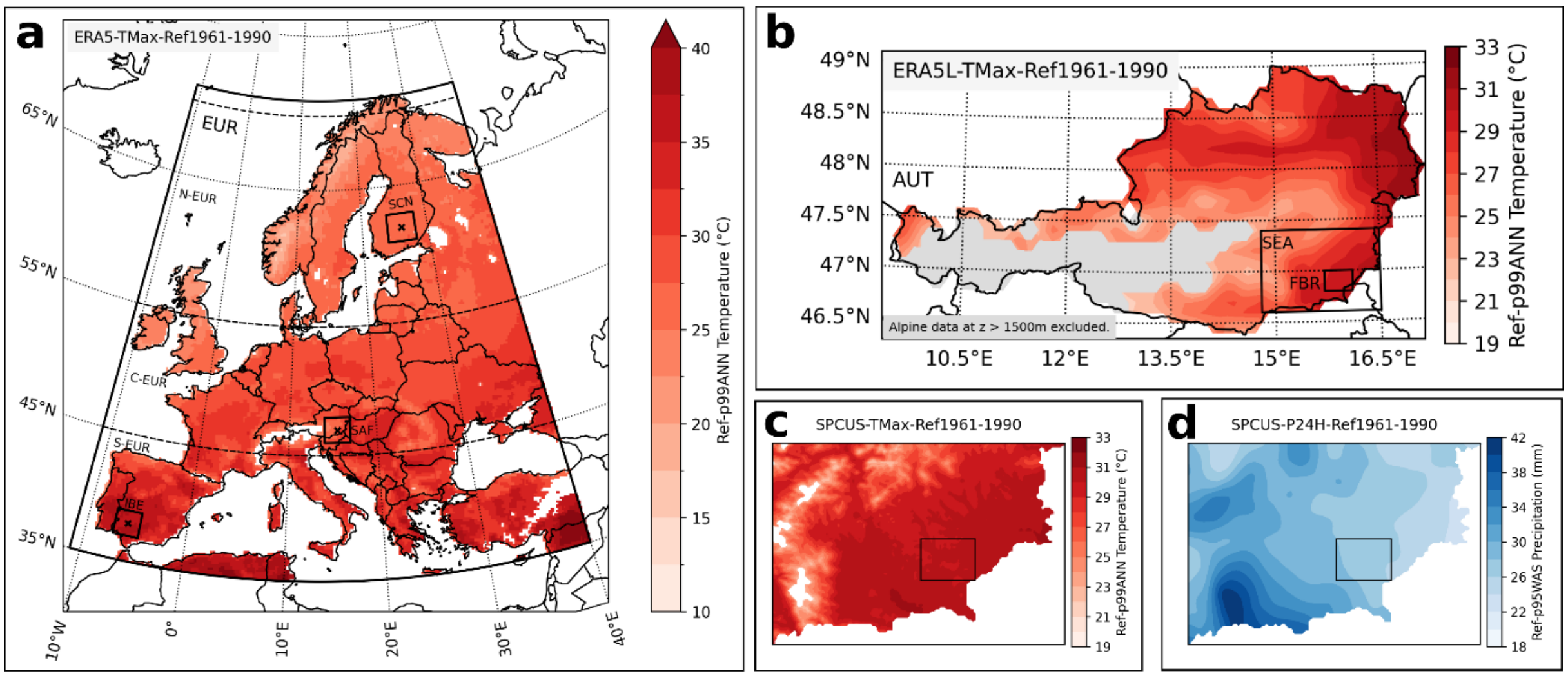

**Extended Data Fig. 1 | Reference threshold maps for the European and Austrian GRs complementary to Fig. 1d,e.** In addition to Fig. 1d,e depicting the 99th-percentile $T^{Max}$ reference threshold maps for AUT and FBR based on the 1-km-scale SPARTACUS data, this figure shows them for the GRs across the whole European land area (EUR) based on the 30-km-scale ERA5 data (**a**), the country GR AUT based on the 10-km-scale ERA5-Land data (**b**), and the sub-country GR SEA based on the 1-km-scale SPARTACUS data (**c**). Furthermore, the SPARTACUS-based 95th-percentile reference threshold map for daily precipitation amounts (of wet days during the warm season April-October) is depicted for the GR SEA (**d**), for which also precipitation extremes are explored through the TEA metrics. Like in Fig. 1d,e, all these maps apply to the Ref1961-1990 period.

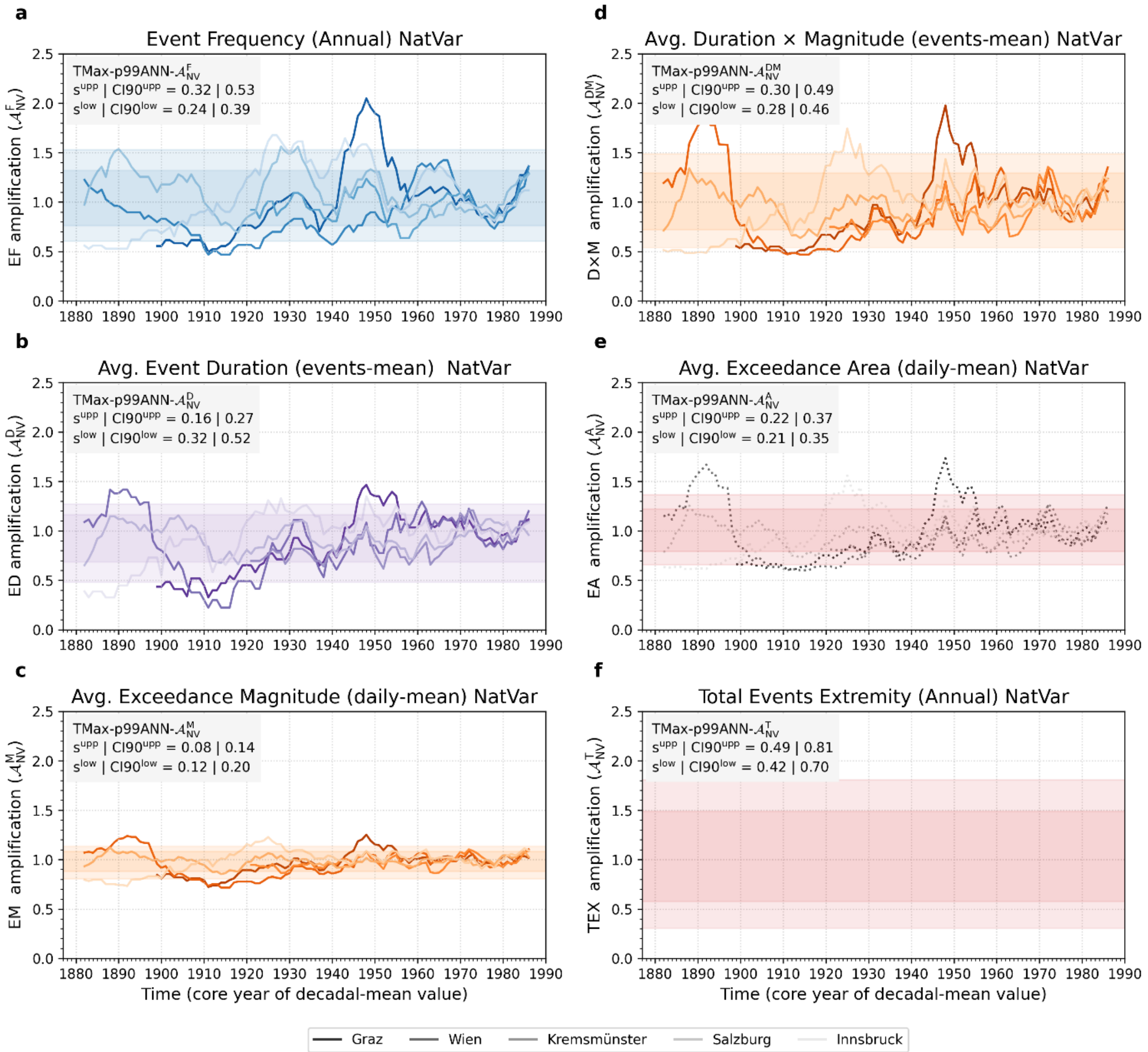

**Extended Data Fig. 2 | Estimated natural variability of amplification factors in the country GR Austria.** Complementing the estimated natural variability corridors shown in Fig. 2 for the GR Austria, this figure illustrates their computation based on the centennial homogenized $T^{Max}$ datasets of five representative stations located across Austria (Graz in the southeast, Vienna in the northeast, Kremsmünster north-central, Salzburg mid-central, Innsbruck in the west) (see Methods for more details). **a-d**, Estimates for the three basic temporal-only metrics frequency (**a**), duration (**b**), magnitude (**c**) and for the related duration × magnitude exceedance metric (**d**). **e**,**f**, Estimates for the spatial-extent-dependent metrics area (**e**) and total events extremity (**f**), which involve a scaling of the temporal duration × magnitude amplification to its well-correlated area amplification. The panels show the natural variability estimates as central (darker-shaded range $1-s^{low}$ to $1+s^{upp}$) and extended (full shaded range $1-CI90^{low}$ to $1+CI90^{upp}$) corridor, with the GR-scaled decadal-mean annual amplification factor time series of the stations overplotted where applicable. The corresponding values of the upper and lower standard deviation ($s^{upp}$, $s^{low}$) and 90% confidence range ($CI90^{upp} = 1.645 \cdot s^{upp}$, $CI90^{low} = 1.645 \cdot s^{low}$) are noted in the legend.

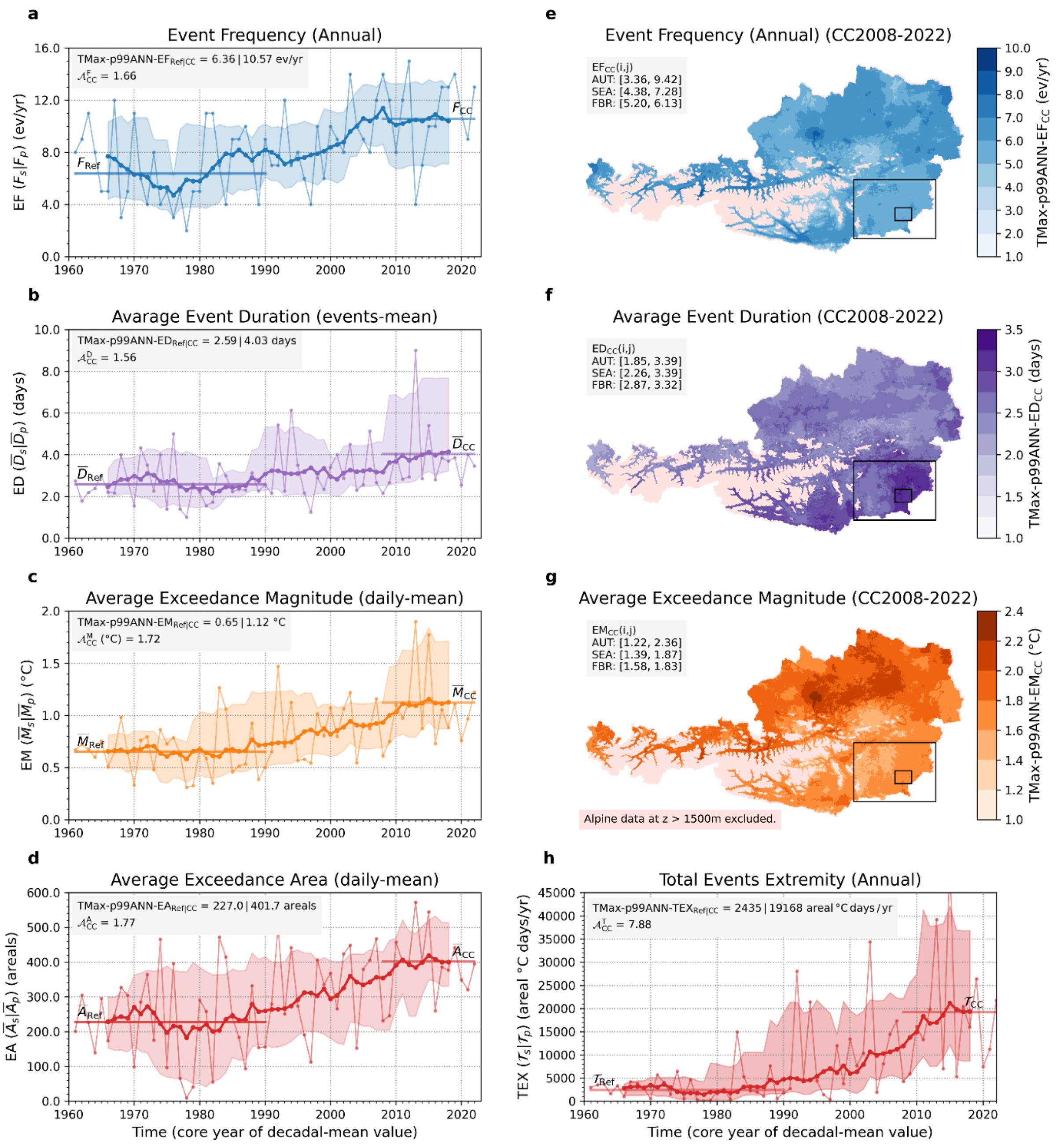

**Extended Data Fig. 3 | Annual and decadal-mean annual TEA metrics of extreme heat events in the country GR Austria.** Complementary to the decadal-mean amplification factor results shown in Fig. 2 for the GR Austria, this figure directly illustrates the native annual and decadal-mean annual metrics in the same setup of panels, from time series of the basic GR metrics frequency (**a**), duration (**b**), magnitude (**c**) and area (custom unit: 1 areal = 100 km$^2$) (**d**) via maps of the temporal local-scale metrics frequency (**e**), duration (**f**) and magnitude (**g**) for the CC2008-2022 period to the time series of the compound TEX (**h**). The time series panels (**a-d**,**h**) show the GR metrics (light connected symbols: annual data 1961-2022, the year-2015 TEX value clipped in panel **h** is ~52000 areal °C days/yr; heavy connected symbols: decadal-mean annual data of core years 1966-2018; horizontal lines: Ref1961-1990 and CC2008-2022 values; legend as Fig. 2); the shaded corridor shows the estimated upper and lower standard deviations of the native annual values (ten per decadal window) about their decadal-mean core year value. Panels (**e-g**) co-list the [Minimum, Maximum] values of all grid points in their legend; the $EF_{CC}(i,j)$ and $ED_{CC}(i,j)$ are smaller than the associated GR metric that collects exceedances across the full GR area, while the $EM_{CC}(i,j)$ are larger as they involve no spatial smoothing.

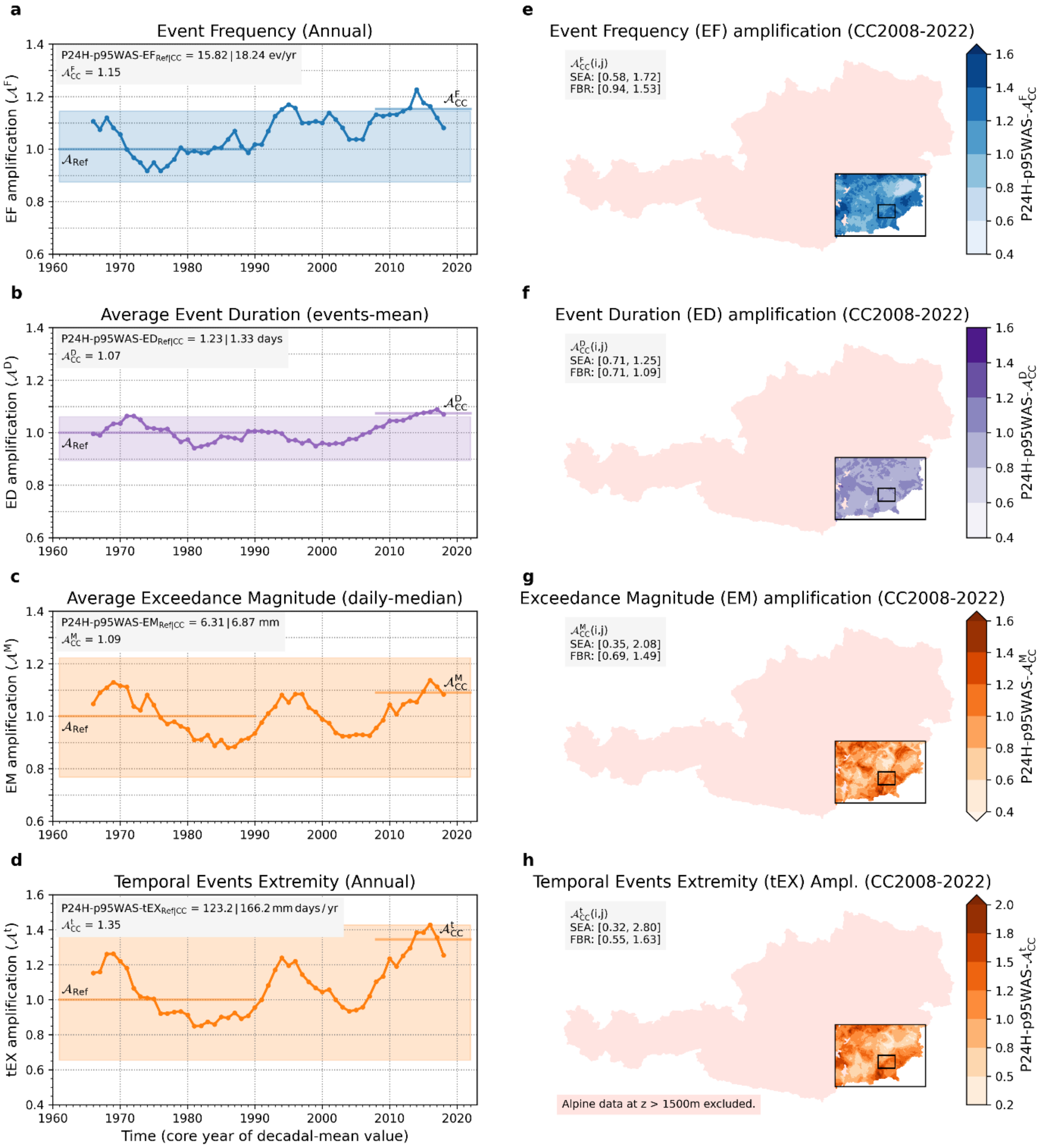

**Extended Data Fig. 4 | Tracking the amplification of extreme precipitation events in the sub-country GR Southeast Austria.** Complementary to the extreme heat amplification results shown in Fig. 2 for the GR AUT (including the sub-country GRs SEA and FBR), this figure depicts the results for precipitation extremes in the GR SEA (including FBR). They are also based on SPARTACUS data (warm-season 24H precipitation amounts), using the threshold map shown in Extended Data Fig. 1d, and displayed in the same panels format. Due to the high spatial variability of the rainfall foreclosing robust exceedance area estimates, the focus here is on the temporal TEA metrics, including time series for the GR metrics frequency (**a**), duration (**b**), magnitude (**c**) and compound tEX (**d**) along with maps of the corresponding local-scale metrics for the CC2008-2022 period (**e-h**). The style and type of content is the same as for Fig. 2 panels; see that caption for further description.

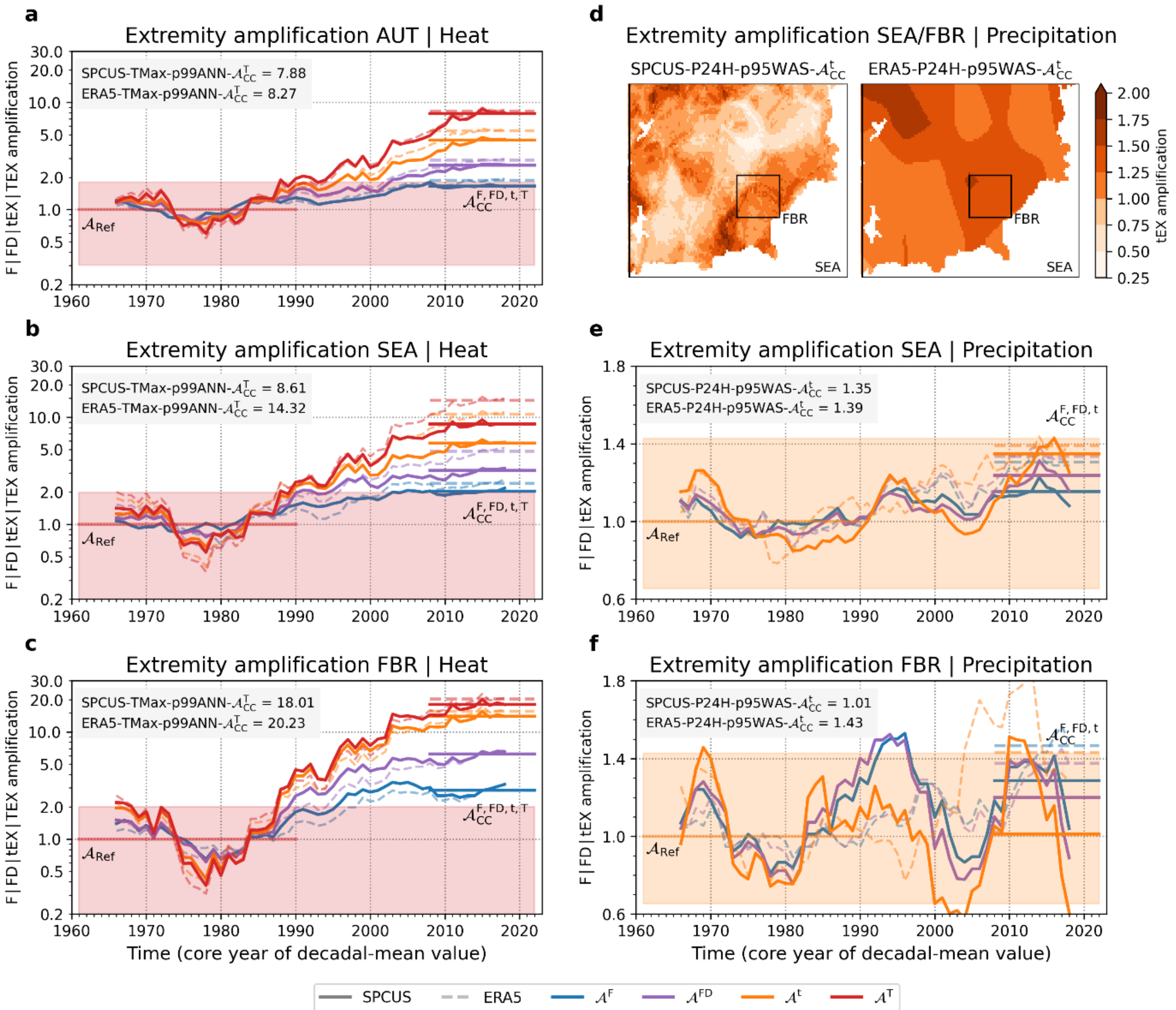

**Extended Data Fig. 5 | Fine-grained insight to extreme heat and precipitation amplification comparing SPARTACUS and ERA5 data in the GRs of Austria.** Complementary to the extreme heat and precipitation amplification results shown in Fig. 3 based on the 1-km-scale SPARTACUS and 10-km-scale ERA5-Land data, this figure compares the same SPARTACUS results to the ones based on the 30-km-scale ERA5 data. The panels layout (**a-c** for heat extremes, **d-f** for precipitation extremes) as well as the style and type of content is the same as for Fig. 3 (with ERA5-Land results replaced by ERA5 results); see that caption for further description.

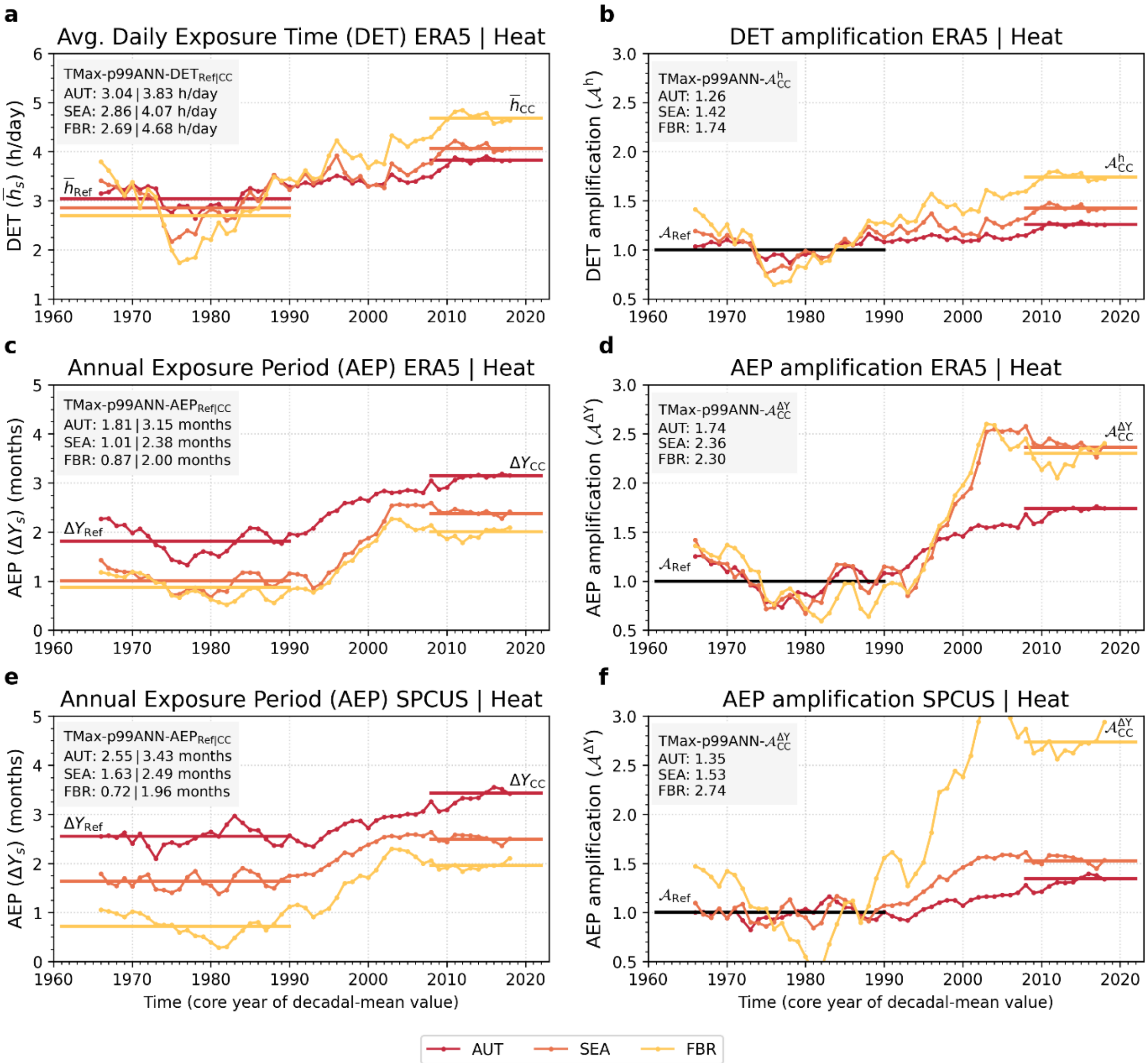

**Extended Data Fig. 6 | Amplification of daily exposure time and annual exposure period due to heat extremes in the example GRs of Austria.** Complementary to the extreme heat amplification results shown in Fig. 3a-c and Extended Data Fig. 5a-c for the GRs AUT, SEA and FBR, this figure shows the supplementary decadal-mean annual DET and AEP metrics and their amplification factor time series (as in Fig. 2a-d,h) based on the 30-km-scale ERA5 (DET and AEP) and 1-km-scale SPARTACUS (AEP) data. **a**,**b**, DET changes along the core years 1966-2018 (**a**) and their amplification vs. Ref1961-1990 (**b**) in the GRs AUT (red), SEA (orange) and FBR (yellow) based on hourly ERA5 data, with numerical values for the Ref and CC periods co-listed in the panel legends. Multiplying the DET with the TEX will yield the total events extremity for hourly heat exposure, hTEX (areal °C h/yr), and the related compound amplification factor $A^{\mathbf{hT}} = A^{\mathrm{h}} \cdot A^{\mathrm{T}}$. **c**-**f**, AEP changes and their amplification for the same GRs based on daily ERA5 (**c**,**d**) and SPARTACUS (**e**,**f**) data, with numerical values for the Ref and CC periods as well co-listed in the panel legends.

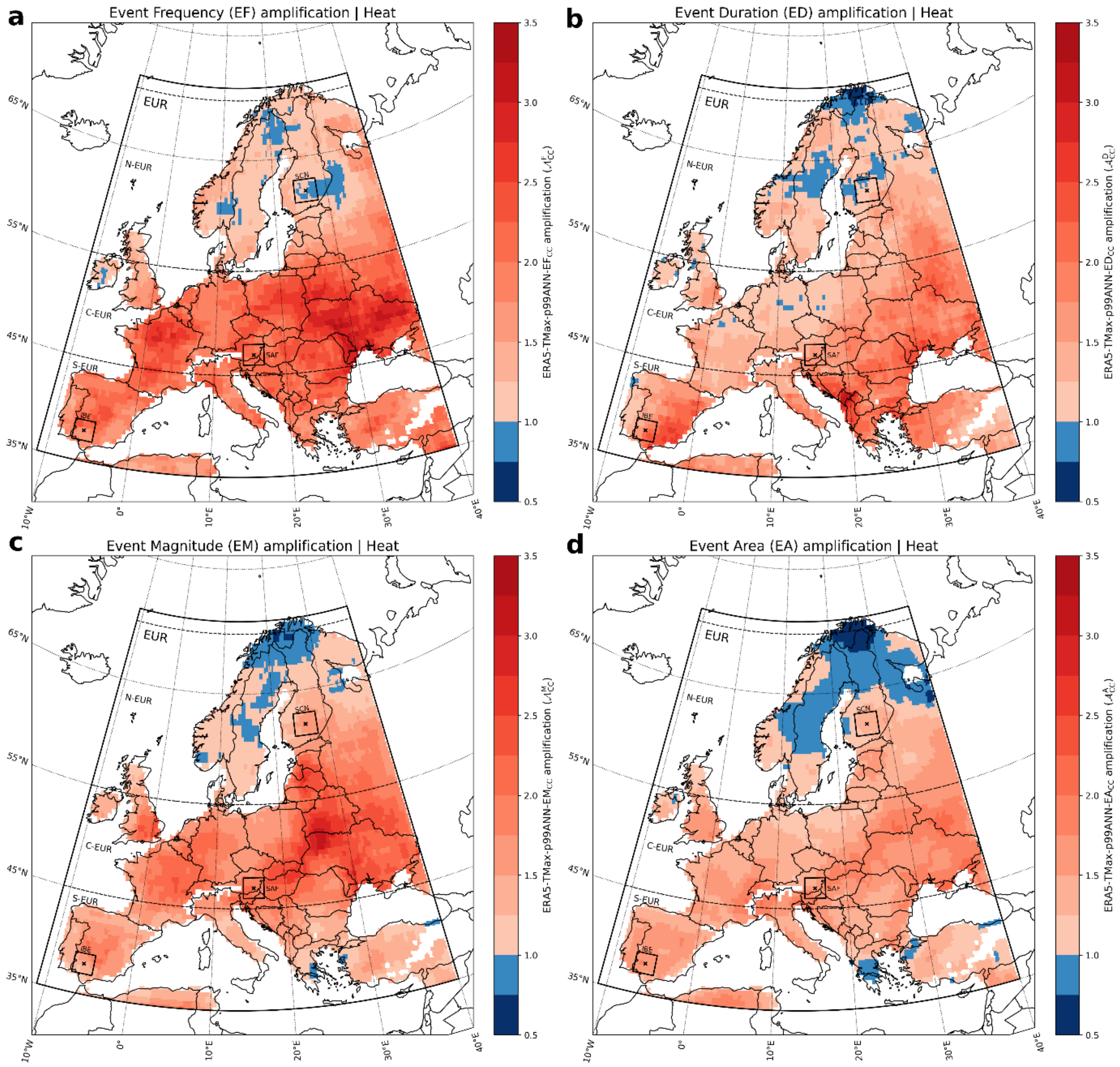

**Extended Data Fig. 7 | Decomposing the TEX amplification of heat extremes over Europe into contributions of frequency, duration, magnitude, and area changes.** Complementary to the map of TEX amplification for CC2008-2022 vs. Ref1961-1990 shown in Fig. 4a for the gridded GRs across Europe, this figure depicts the corresponding results of the decomposition into the individual TEA metrics of event frequency (**a**), event duration (**b**), exceedance magnitude (**c**) and exceedance area (**d**) amplification ($A^T = A^F \cdot A^D \cdot A^M \cdot A^A$). The style and type of content of panels **a-d** is the same as in Fig. 4a (with the TEX results replaced by the ones for the basic metrics and the value range of the color bar reduced); see that caption for further description.

# Methods

## Scope and applicability of the TEA metrics

Generically framed, the threshold-exceedance-amount (TEA) indicators are a new class of self-consistent extremity metrics, which are derived through a customized space-time filter for cascaded and multi-scale extraction of extremes and their statistics along time in threshold-exceedance subspaces of a spatiotemporal field of a key variable of interest, conditioned on a space-domain-threshold map intersecting the key variable. The threshold exceedance amounts (TEAs) are obtained along time at the TEA sampling rate chosen (e.g., annually), based on key variable data at a suitable native time resolution (e.g., daily). They form a multiplicative class of basic statistics of "exceedance events", including event frequency, mean event duration, and mean magnitude and spatial extent of exceedance as the main ones.

The key variables need to be metric-scaled scalar or vector quantities (rather than categorical variables) and available in form of spatiotemporal fields of adequate accuracy, resolution and coverage of the application domain and space-time scales of interest, such as in this study in form of long-term daily gridded fields of climate-quality temperature data across Europe for tracking heat extremes. The space dimension of the fields can range from single-location (e.g., station data) via one-dimensional (e.g., altitude-dependent) and two-dimensional (as in this study) to three-dimensional (e.g., latitude-longitude-altitude data cubes).

Formally also key variables in parameter spaces, including higher dimensional ones, rather than in real space (like the near-Earth-surface space in this study) could be explored for TEAs. Therefore, given the very generic scope of this threshold-based space-time filtering as a data analytics method, applications also far beyond weather and climate extremes in Earth's climate system—for example, investigations of changes in extremes in other physical or in socio-economic or biological systems—could benefit from employing the new metrics class.

For pertinence for the study and ease of uptake, but without loss of the general applicability of the methodological description, we here introduce the TEA metrics along the study focus of exploring heat and partly precipitation extremes, from local- to country-scale (in the example country Austria) up to aggregate continental-scale regions of Europe; for illustration see Fig. 1 and Extended Data Fig. 1, for a detailed algorithmic description the Supplementary Note.

Based on multi-decadal daily gridded datasets of key climate variables (daily maximum and hourly temperatures and daily precipitation amounts in this study), the scope of the metrics for tracking extreme weather and climate events under climate change is that they:

(1) rigorously and consistently quantify local-to-regional annual and decadal-mean changes in event frequency (EF), event duration (ED), exceedance magnitude (EM) and exceedance area (EA)—optionally also in timing characteristics including daily exposure time (DET) and time-of-day shifts in exposure, and annual exposure period (AEP) and day-of-year shifts of first and last events—as separate metrics, partially compound as temporal events extremity (tEX) and event severity (ES), and as total events extremity (TEX) and optionally hourly (heat) exposure TEX (hTEX = DET × TEX);

(2) flexibly focus on any key-variable-data-covered georegions (GRs; from 10-km-scale to 100-km-scale/whole-countries size) and any annual climatic time periods of interest (CTPs; from full year (ANN) or warm season April-October (WAS) via extended summer May-September or extended winter November-March to any season or month), providing local-scale gridded results at the input data grid within GRs as well as whole-GR metrics;

(3) are based on a grid-resolved reference threshold map in the key climate variable that delineates grid values considered extreme from those considered regular, employing a high-percentile-based threshold (e.g., p99 values computed from the data in a reference period 1961-1990) or any other suitable threshold formulation, and deliver the class of metrics in a strictly cascaded manner up to TEX (and optionally hTEX) as the overall metric;
(4) also include decadal-mean annual amplification factors over the recent decades vs. a suitable reference period (e.g., 1961-1990 as being not much influenced yet by climate change) for all metrics and—subject to availability of proper auxiliary multi-centennial pre-1990 data—associated natural variability estimates that can be used to support climate change attribution of anthropogenic shares in the amplification;
(5) and include aggregate-georegion (AGR) metrics, derived as means along with spread estimates from consistent spatial averaging over the individual GR metrics within an AGR, for tracking extremes over any larger-scale regions of interest, from countries to continental-scale assessment regions (e.g., the Europe-wide land regions used in this study).

**Georegions and datasets**

Georegions (GRs) are study regions of interest, of a few 100 km$^2$ size up to whole-countries size or ~50,000 km$^2$ geographic-cell size, wherein the potential amplification of extremes due to climate change is tracked by the TEA indicators. For this study, a cascaded multiple-size-scales ensemble of exemplary GRs in Europe was selected (see Fig. 1c-d), which comprises:
(1) Austria as a country-region example (GR AUT; country area 83,883 km$^2$, with 66,988 km$^2$ at ≤ 1,500 m altitude counted into the GR);
(2) its regional-scale subregion Southeast Austria (GR SEA; area 9,795 km$^2$ in the UTM[58] zone 33N 5,157,000–5,249,000N × 495,000–624,000E within Austria, with 9,622 km$^2$ at ≤ 1,500 m altitude counted into the GR);
(3) the Feldbach Region embedded within SEA as local-region example of smallest size (GR FBR; 414 km$^2$, UTM zone 33N 5,188,000–5,206,000N × 557,000–580,000E);
(4) and, as a continent-wide area covered by GRs, a whole European land grid of GRs (0.5° × 0.5° latitude-longitude grid within 34°N–72°N × 12°W–41°E; GR grid area around each grid point ±1° × ±1°/cos(latitude), size 49,457 km$^2$; grid areas included as eligible GR if covered by at least 50 % land after excluding subareas covered by sea or exceeding 1500 m altitude).

The aggregate georegions (AGRs) were defined based on the European grid of GRs and include four continental-scale AGRs (Fig. 1c):
(1) whole European land (EUR; all eligible grid points in 35.0°N–70.0°N × 11.0°W–40.0°E, total area 7,666,936 km$^2$);
(2) Central Europe land (C-EUR; all eligible grid points in 45.0°N–55.0°N × 11.0°W–40.0E°N, total area 3,252,386 km$^2$);
(3) Southern Europe land (S-EUR; all eligible grid points in 35.0°N–44.5°N × 11.0°W–40.0°E, total area 2,241,613 km$^2$); and
(4) Northern Europe land (N-EUR; all eligible grid points in 55.5°N–70.0°N × 11.0°W–40.0°E, total area 2,172,937 km$^2$).
Within each of the three AGRs C-EUR, S-EUR and N-EUR also a representative individual GR was defined for closer inspection (see Fig. 1c and Fig. 4a-b), including:
(1) the Southeastern Alpine Forelands GR (SAF; center grid point 47.0°N, 15.5°E);
(2) an Iberian GR (IBE; center 38.0°N, 6.0°W); and
(3) a Scandinavian GR (SCN; center 62.0°N, 26.0°E).

The input datasets, required area-wide in a GR to compute the TEA metrics for a weather or climate extreme (climate hazard), comprise the daily gridded data of the key variable adopted to characterize the hazard together with the related gridded data for the grid cell areas and threshold values (see Reference threshold maps). Furthermore, if available, supplementary sub-daily input data at hourly resolution can be used to additionally characterize the sub-daily exposure to the hazard (e.g., by computing the DET metric), and auxiliary (multi-)centennial pre-1990 data (e.g., from historic observations or climate model control runs) can be used to compute natural variability estimates associated with the amplification factor metrics.

In this study we employed the following input datasets for the key variables daily maximum temperature ($T^{\mathrm{Max}}$; CTP = ANN), hourly temperatures ($T^{m}$, $m$=1,…,24 h per day; CTP = ANN) and daily precipitation amount ($P^{\mathrm{24H}}$; CTP = WAS) for the selected GRs:

(1) 1-km-scale Austrian SPARTACUS v1.5 data[28,29] (SPCUS; 1 km × 1 km grid in GRs AUT, SEA, FBR; $T^{\mathrm{Max}}$, $P^{\mathrm{24H}}$ over 1961-2022; v1.5 used instead of newer v2.1, since $T^{\mathrm{Max}}$ is very similar but $P^{\mathrm{24H}}$ is resolution-degraded due to less stations in v2.1);

(2) 10-km-scale European Reanalysis 5 Land data[30] (ERA5L; 0.1° × 0.1° grid in GRs AUT, SEA, FBR; $T^{\mathrm{Max}}$, $T^{m}$, $P^{\mathrm{24H}}$ over 1961-2022);

(3) 30-km-scale European Reanalysis 5 data[31] (ERA5; 0.25° × 0.25° grid covering all EUR GRs and the GRs AUT, SEA, FBR; $T^{\mathrm{Max}}$, $T^{m}$, $P^{\mathrm{24H}}$ over 1961-2022).

Data grid cells not classified land or exceeding 1500 m altitude are excluded (based on the surface altitude data being part of each dataset) and cell areas are assigned according to the grid resolution, with cell areas overlapping GR boundaries clipped to the GR-internal fractional area.

For cross-check of the ERA5-based DET results for the Austrian GRs and extreme heat amplification results for the gridded European GRs as part of extensive sensitivity tests, we in addition used hourly ERA5-Land $T^{m}$ data and daily 0.25° × 0.25°-gridded E-OBS[59] v29.0e $T^{\mathrm{Max}}$ data, respectively; both led to reasonably similar results in the cross-checked metrics, consistent with expected uncertainties due to differences in the datasets involved. For computing the ERA5-based atmospheric heat content gain (AHCg; see Fig. 4d), the same algorithms as used in ref. 32 for AHCg computation were applied here.

For estimating the natural variability of decadal-mean amplification factors in the GRs AUT and SEA, we used long-term historic time series of $T^{\mathrm{Max}}$ and $P^{\mathrm{24H}}$ up to 1990 from overall seven representative stations with daily data since the late 19th century (GeoSphere Austria Daily data v2; see Data availability). These were crosschecked for long-term homogeneity by monthly HISTALP station data[40] ($T_{\mathrm{MonMean}}$, $P_{\mathrm{MonTotal}}$) and underpinned for the heat extremes by further backward assessment based on multi-centennial ModE-RA monthly reanalysis data[34,60] ($T_{\mathrm{MonMean}}$ 1422-1990) over Austria (see Supplementary Note, section 6.3 therein, for details).

While global-coverage input datasets like the ERA5 reanalysis (and many further datasets) enable to equally explore any other GRs worldwide and many other types of extreme events (e.g., ref. 2,17–19) for new levels of insight, we restricted to these European GRs and datasets and a focus on extreme heat here, in the spirit of a compact pioneer study breaking ground for the new class of metrics but leaving further applications to follow-on studies.

### Reference threshold maps

The spatial reference threshold maps, required on the input data grid within a GR along with the key variable fields, serve to delineate at any grid point key variable values along time that are considered extreme from those considered regular, with the threshold values for a key variable defined in a "reference climate"[2,3]. For this study, we computed the threshold values for each selected GR and input dataset (see Georegions and datasets) as a map of suitable high-percentile values based on the respective daily gridded data within 1961-1990, which we adopted as the reference climate period being not much influenced yet by climate change.

Specifically we computed, at every grid point, the 99th percentile (p99) value of daily maximum temperature ($T^{\mathrm{Max}}$) of all days-of-year (ANN) over 1961-1990 (Ref) for delineating heat extremes ("TMax Ref-p99ANN Temperature"; threshold map results illustrated in Fig. 1d-e and Extended Data Fig. 1a-c), and the 95th percentile (p95) value of daily precipitation amount ($P^{24\mathrm{H}}$) of all wet ($P^{24\mathrm{H}} \geq 1$ mm) days-of-year during the warm season April-October (WAS) for delineating precipitation extremes ("P24H Ref-p95WAS Precipitation"; GR SEA SPCUS results illustrated in Extended Data Fig. 1d). For the latter, given the high spatial variability of precipitation in the SEA region[37–39] that is typical for mid-latitude regions with a dominance of local-scale convective rainfall during summertime[37,61,62], a smoothing over an area of ~7 km radius around each grid point was found suitable and hence used for obtaining the final GR SEA SPCUS threshold values.

We carefully tested the sensitivity of the results to the choice of threshold maps and found all essential outcomes of the study robust, and the metrics showing plausible quantitative changes to percentile variations within p90-p99.8 along with other reasonable choices (like CTP ANN, WAS or ESS/extended summer season May-Sep; Ref 1951-1980, 1961-1990 or 1971-2000). As seen from the threshold maps in Fig. 1 and Extended Data Fig. 1, a clear advantage of using a percentile-based threshold formulation is that it can sensibly reflect local-scale climate conditions, such as topographic effects and the latitudinal gradient in the $T^{\mathrm{Max}}$ threshold in the case of heat extremes, neatly delineating what is extreme for populations and ecosystems at a specific location or in a specific region.

It is nevertheless important for any use to understand the sensitivity of results to the choice of threshold map (e.g., the percentile chosen), since this map fundamentally splits what is chosen as "extreme", the desired TEA subspace, from the much larger "regular" subspace that is discarded. 30-yr percentiles > p99.8, corresponding to a 10-yr return period or rarer, are not recommended; for such very rare extremes the use of extreme-value-theory (EVT) methods[14,53,54] will generally be a more suitable choice.

From a general perspective, the choice of threshold formulation for computing the TEA metrics will depend on application, that is on its fitness-for-purpose, and evidently a large number of threshold-defined indices for a wide diversity of purposes already exists[2,12–20]. In general, therefore, any other threshold formulation suitable for an application of interest can be used as well, such as simple constant values (like for hot days ≥ 30 °C) or day-of-year dependent thresholds (for expressing seasonal variations) but also upper-bound thresholds instead of lower-bound ones as used here (for low-value exceedances like by cold extremes) or even range-confining thresholds bounding TEAs from above and below (e.g., for tipping-point-relevant tracking of changes in a critical control parameter value range[63,64], like of temperature changes around the 0 °C ice-to-water phase transition).

**Computation of the metrics**

Building on (1) the definition of georegions and the selection of associated daily gridded key variable input datasets (see Georegions and datasets), (2) the reference threshold maps derived based on the key variable data or defined otherwise (see Reference threshold maps), and (3) a detailed scientific-technical definition of spatial and temporal sampling indices and of all variables involved (see Supplementary Note, section 2 therein on Index and variable definitions), the TEA indicators are computed along the following five-steps sequence.

This is done on the stacked-vector input data grid $k(i,j)$ of eligible grid cells per GR (cells classified land and ≤ 1500 m) as well as for whole GRs. The inputs per GR include the daily ($n$) key variable data $x_{nk} = x_n(i,j)$ [$T_n^{\mathrm{Max}}(i,j)$ or $P_n^{24\mathrm{H}}(i,j)$] for $n$=1-22645 days over 1-Jan-1961 to 31-Dec-2022, optionally also $m$=1-24 hourly data each day $x^m_{nk} = x^m_n(i,j)$ [$T^m_n(i,j)$ or $P^m_n(i,j)$];

the threshold map $x_k^{\mathrm{Thres}} = x^{\mathrm{Thres}}(i,j)$ [$T^{\mathrm{Thres}}(i,j)$ or $P^{\mathrm{Thres}}(i,j)$]; the grid cell areas $a_k = a(i,j)$ and related total area $A^{\mathrm{GR}}$; and the minimum daily threshold exceedance area $A^{\mathrm{min}}$ required for counting a day $n$ as GR exceedance day (set to 1 areal = 100 km$^2$). This is complemented in the final aggregate-georegions step by building on a stacked-vector grid $l(i,j)$ of eligible GR grid cells per AGR (GR cells with > 50 % land at ≤ 1500 m), their cell areas $a_l$ and related total area $A^{\mathrm{AGR}}$, and the GR decadal-mean results for the TEA metrics on the $l(i,j)$ grid.

**Computation of daily basis variables.**

The main daily basis variables, computed after first initializing all of their elements with 0, comprise:

(1) the daily threshold exceedance count $c_n \in \{c_{nk}, c_n^{\mathrm{GR}}\}$ (basic algorithm: if $x_n$ exceeds $x^{\mathrm{Thres}}$ then $c_n$=1, for $c_n^{\mathrm{GR}}$=1 also exceedance area ≥ $A^{\mathrm{min}}$ needed);

(2) the daily threshold exceedance event count $e_n \in \{e_{nk}, e_n^{\mathrm{GR}}\}$ (basic algorithm: for each cluster of consecutive $c_n$=1 days set the core day $e_n$=1);

(3) the daily threshold exceedance magnitude $m_n \in \{m_{nk}, m_n^{\mathrm{GR}}\}$ (basic algorithm: if $c_n$=1 then $m_n = x_n - x^{\mathrm{Thres}}$); and

(4) the daily threshold exceedance area $a_n^{\mathrm{GR}}$ (basic algorithm: if $c_n^{\mathrm{GR}}$=1 then $a_n^{\mathrm{GR}} = \Sigma_k a_k | c_{nk}$=1).

The count and magnitude metrics are evidently computed both at the local-scale input data grid $k(i,j)$ and for the whole GR, while the area can only be computed for the GR.

The main supplementary daily basis variable, computed in case hourly data $x^m{}_{nk}$ are available, is the daily number of exceedance hours $N^{\mathrm{h}}{}_n \in \{N^{\mathrm{h}}{}_{nk}, N^{\mathrm{h}}{}_n{}^{\mathrm{GR}}\}$ (basic algorithm: if $c_n$=1 then $N^{\mathrm{h}}{}_n$ = Count($m$) | $x^m{}_n > x^{\mathrm{Thres}}$). For the detailed algorithms see Supplementary Note, section 3 therein.

These metrics are typical quantities also widely used in one or the other variant for separate computation of many conventional threshold-defined indices such as hot days, event frequency or exceedance intensities[2,12,13,15]. Most akin to this TEA concept, a recent study used a preconception of the metrics with a focus on exploring long-term changes in daily threshold exceedance areas of temperature anomalies in different layers of the stratosphere to monitor sudden stratospheric warmings under climate change based on ERA5 reanalysis data[20].

However, while providing valuable finest-resolution TEA insights on their own for tracking details of individual events, these daily variables in particular serve as input to the subsequent computation of the annual indicator variables.

**Computation of annual indicator variables.**

Building on the daily metrics over the 62 years ($p$) from $p$=1961 to $p$=2022 in the chosen annual climatic time period (CTP days $q$ of all days $n$ of year $p$; all days if CTP = ANN, April-October days if CTP = WAS), we first compute the four basic annual TEA variables in a cascaded sequence (with all of their elements initialized with 0 beforehand), including:

(1) the annual event frequency (EF), $F_p \in \{F_{pk}, F_p^{\mathrm{GR}}\}$, expressing the number of events in the annual CTP (unit: 1/yr = yr$^{-1}$, verbally also events/year (in the CTP); basic algorithm: if any $e_q > 0$ in year $p$ then $F_p = \Sigma_q e_q$);

(2) the average event duration (ED), $\overline{D}_p \in \{\overline{D}_{pk}, \overline{D}_p^{\mathrm{GR}}\}$, expressing the events-mean duration of all events $\Sigma_q e_q = F_p$ in the CTP (unit: days, verbally also days/event; basic algorithm: if $F_p > 0$ then $\overline{D}_p = \Sigma_q c_q / F_p$);

(3) the average exceedance magnitude (EM), $\overline{M}_p \in \{\overline{M}_{pk}, \overline{M}_p^{\mathrm{GR}}\}$, expressing the daily-mean EM (for $T^{\mathrm{Max}}$) or daily-median EM (for $P^{\mathrm{24H}}$) of all exceedance days $\Sigma_q c_q = D_p = F_p \cdot \overline{D}_p$ in the CTP (unit: °C or mm; basic algorithm: if $F_p > 0$ then $\overline{M}_p = \Sigma_q m_q / D_p$); and

(4) the average exceedance area (EA) in the GR, $\bar{A}_p = \bar{A}_p^{GR}$, expressing the EM-weighted daily-mean EA over all daily GR exceedance magnitudes $\Sigma_q m_q^{GR} = M_p^{GR} = D_p^{GR} \cdot \bar{M}_p^{GR}$ in the CTP (unit: areals, 1 areal = 100 km$^2$, a custom unit matching the scales of interest in space, similar to the unit days in time; basic algorithm: if $F_p^{GR} > 0$ then $\bar{A}_p = \Sigma_q m_q^{GR} a_q^{GR} / M_p^{GR}$).

Based on this self-consistent class of basic annual TEA metrics $X_p \in \{F_p, \bar{D}_p, \bar{M}_p, \bar{A}_p\}$, the compound metrics are obtained by simple multiplication, including as the three main ones:

(1) the temporal events extremity (tEX), $M_p \in \{M_{pk}, M_p^{GR}\}$, expressing the temporal CTP-cumulative EM $\Sigma_q m_q = M_p = F_p \cdot \bar{D}_p \cdot \bar{M}_p$ at the local-scale grid or for the GR (unit: °C days/yr or mm days/yr);

(2) the average event severity (ES) in the GR, $\bar{S}_p = \bar{S}_p^{GR} = \bar{D}_p^{GR} \cdot \bar{M}_p^{GR} \cdot \bar{A}_p^{GR}$, expressing the CTP-mean overall exceedance strength per event (unit: areal °C days or areal mm days); and

(3) the total events extremity (TEX) in the GR, $\mathcal{T}_p = \mathcal{T}_p^{GR} = F_p^{GR} \cdot \bar{S}_p^{GR}$, expressing the spatiotemporal CTP-cumulative EM (in areal °C days/yr or areal mm days/yr).

Any intermediate “twin products”, such as the CTP-cumulative events duration $D_p = F_p \cdot \bar{D}_p$ (days/yr), can be of application-specific interest as a metric as well. For example, if using a $T^{Max}$ threshold of 35 °C (conventionally termed TX35 [12]), $D_p$ would express the number of extremely hot days in the chosen annual CTP that are critical for maize pollination and production in agriculture[12,65,66] and particularly hazardous for human health[3,11,12].

The main supplementary annual indicator variables computed in addition (with all of their elements first initialized with 0) comprise the average daily exposure time (DET), $\bar{h}_p \in \{\bar{h}_{pk}, \bar{h}_p^{GR}\}$, expressing the average number of exceedance hours per day of all exceedance days $D_p$ in the CTP (unit: h/day; basic algorithm: if $F_p > 0$ then $\bar{h}_p = \mathrm{Avg}(N^h{}_q)$ over all $c_q$=1 days), and the annual exposure period (AEP), $\Delta Y_p \in \{\Delta Y_{pk}, \Delta Y_p^{GR}\}$, expressing the difference between the day-of-year (DOY) of the first and last events in the CTP (unit: months; basic algorithm: if $F_p > 0$ then $\Delta Y_p = \{\mathrm{DOY}[\mathrm{Max}(q|e_q{=}1)] - \mathrm{DOY}[\mathrm{Min}(q|e_q{=}1)] + 1\}/30.5$). Combining DET and TEX yields the total hourly (heat) exposure hTEX, $\bar{h}_p^{GR} \cdot \mathcal{T}_p^{GR}$ (unit: areal °C h/yr).

Furthermore, for specifically inspecting the near-surface extreme heat energy increase (see main text section Amplification of heat extremes and atmospheric heat over Europe), we estimated the threshold exceedance energy deposited annually in the atmospheric boundary layer (ABL) of a GR by way of the ABL exceedance heat content (AEHC), $H_{\mathrm{AEHC},p} = H^{GR}_{\mathrm{AEHC},p} = \tilde{C}_{ABL} \cdot \mathcal{T}_p^{GR}$. This compound metric expresses the CTP-cumulative AEHC proportional to the TEX, using an approximative ABL excess heat uptake capacity $\tilde{C}_{ABL}$ = 1.507 MJ m$^{-2}$ °C$^{-1}$ day$^{-1}$ = 0.1507 PJ/(areal °C day) as the conversion factor derived for typical mean ABL conditions during summertime over Europe[67–69]. For the detailed algorithms for all annual indicator variables see Supplementary Note, section 4 therein.

These annual TEA metrics, for which Extended Data Fig. 3 includes an illustration as part of the study of heat extremes in the country GR Austria, provide a rich resource for fine-grained insight into the year-to-year evolution of extremes and many of them are, in separated form, also the basis for conventional threshold-defined indices[2,12,13,15].

Beyond the above-mentioned sudden-stratospheric-warmings study[20], specific metric definitions closest to the overall compound metrics of the new holistic class (tEX, TEX, hTEX) have been recently used under the notion of heatwave cumulative intensity in studies on heatwave trends over Europe related to jet stream states[36] and on worldwide trends in regional heatwaves[16]. For tracking long-term changes, it is essential to next use these metrics to derive decadal-mean indicators.

**Computation of decadal-mean indicator variables.**

Using first the basic annual metrics of non-compound nature (main ones $X_p \in \{F_p, \overline{D}_p, \overline{M}_p, \bar{A}_p; \bar{h}_p, \Delta Y_p\}$), we derive the corresponding decadal-mean annual time series $X_s$ by moving-window averaging along the 53 decade-window core years ($s$) from $s$=1966 to $s$=2018 (algorithm: $X_s$ = Avg($X_p$) over all years $p\,|\,s–5 \leq p \leq s+4$ of the decade-window about core year $s$).

Subsequently, the compound decadal-mean metrics (main ones $X_s \in \{M_s, \bar{S}_s, \mathcal{T}_s; H_{\mathrm{AEHC},s}\}$) are computed for each core year by multiplication of the respective basic metrics $X_s$ they are composed of. In addition, for indicating the spread of the ten individual-year values $X_p$ about the decadal-mean value $X_s$ of any core year $s$, we co-compute upper and lower standard deviation estimates ($s^{\mathrm{upp}}_{\mathrm{X},s}$, $s^{\mathrm{low}}_{\mathrm{X},s}$) as complementary spread estimator time series for all metrics (Extended Data Fig. 3 includes an example illustration of these spread estimates, for heat extremes in the GR Austria).

On top of the decadal-mean time series $X_s$, we compute the two longer-term means $X_{\mathrm{Ref}}$ and $X_{\mathrm{CC}}$; the former, as for the threshold maps, for the reference period Ref1961-1990 that is not much influenced yet by climate change, the latter for a most recent "current climate" period 2008-2022 (CC2008-2022), chosen to express the present climate as closely as possible while still keeping a reasonably long 15-year average. Specifically, the $X_s$ data are averaged over the core years $s$ within 1961-1990 (Ref) and 2008-2022 (CC) for which the full decadal data windows fall into the period.

In order to conserve the multiplicative character intrinsic by definition in almost all metrics $X_s$, and to keep the resulting $X_{\mathrm{Ref}}$ and $X_{\mathrm{CC}}$ metrics strictly consistent irrespective of whether time-averaging or variables-multiplication is run first, we perform this averaging as a geometric rather than arithmetic averaging (i.e., averaging over relative rather than absolute deviations). While the difference of the averaging results is found very small (since the $X_s$ values generally lie within a factor of 0.5 to 2 about the mean), the compound metrics like TEX react sensitively and rigorous consistency is hence essential. For algorithm details on the decadal-mean variables see Supplementary Note, section 5 therein.

We note that it is convenient to implement the geometric averaging as arithmetic averaging in log-space, since this is the space, where the multiplicative basic metrics (primary ones $X_s \in \{F_s, \overline{D}_s, \overline{M}_s, \bar{A}_s\}$, and likewise for $X_{\mathrm{Ref}}$ and $X_{\mathrm{CC}}$) can just be added up rather than multiplied and where their fractional contribution to compound metrics (primary ones $X_s \in \{M_s, \bar{S}_s, \mathcal{T}_s\}$, and likewise for $X_{\mathrm{Ref}}$ and $X_{\mathrm{CC}}$) can be properly visualized independent of order of multiplication. For example, Fig. 3 includes illustrations in log-space (i.e., with a logarithmic y-axis). We adopt the decadal-mean annual variables as the major metrics towards exploring long-term changes, since their decade-scale "lens" effectively filters across the volatile annual data while still tracking the longer-term time dynamics well (e.g., see Extended Data Fig. 3).

Turning to the next step of computing amplifications, the longer-term means $X_{\mathrm{Ref}}$ and $X_{\mathrm{CC}}$ are key to help quantify the overall climate hazards amplification that may have emerged since the reference period up to the present ("CC vs. Ref"). Furthermore, the reference period helps to co-estimate the range of natural variability against which the recent amplification may possibly be detected as being caused by anthropogenic climate change.

**Computation of decadal-mean amplification factor variables.**

Normalizing first any basic (non-compound) decadal-mean annual time series (main ones $X_s \in \{F_s, \overline{D}_s, \overline{M}_s, \bar{A}_s; \bar{h}_s, \Delta Y_s\}$) and "current climate" mean value (main ones $X_{\mathrm{CC}} \in \{F_{\mathrm{CC}}, \overline{D}_{\mathrm{CC}}, \overline{M}_{\mathrm{CC}}, \bar{A}_{\mathrm{CC}}; \bar{h}_{\mathrm{CC}}, \Delta Y_{\mathrm{CC}}\}$) by the respective reference-period value $X_{\mathrm{Ref}}$, we obtain the corresponding (dimensionless) amplification factor variables $A^{\mathrm{X}}{}_s = X_s\,/\,X_{\mathrm{Ref}}$ (main ones

$A^{X}{}_{s} \in \{A^{F}{}_{s}, A^{D}{}_{s}, A^{M}{}_{s}, A^{A}{}_{s}; A^{h}{}_{s}, A^{\Delta Y}{}_{s}\}$) and $A^{X}{}_{CC} = X_{CC}/X_{Ref}$ (main ones $A^{X}{}_{CC} \in \{A^{F}{}_{CC}, A^{D}{}_{CC}, A^{M}{}_{CC}, A^{A}{}_{CC}; A^{h}{}_{CC}, A^{\Delta Y}{}_{CC}\}$), respectively. The reference-period amplification factor $A^{X}{}_{Ref}$ is by construction always unity under this normalization ($A^{X}{}_{Ref} = X_{Ref}/X_{Ref} = 1$) and hence serves as the fundamental reference for tracking changes in amplification over time.

Subsequently, the compound amplification factors (main ones $A^{X}{}_{s} \in \{A^{t}{}_{s}, A^{S}{}_{s}, A^{T}{}_{s}; A^{HC}{}_{s}\}$ related to $X_s \in \{M_s, \bar{S}_s, \mathcal{T}_s; H_{AEHC,s}\}$, and likewise for the compound $A^{X}{}_{CC}$) are computed for the decadal-mean annual time series and the CC period by multiplication of the respective basic amplification factors they are composed of. For the detailed algorithms for the decadal-mean amplification factor variables see Supplementary Note, section 6 therein.

Since the compound $A^{X}{}_{s}$ and $A^{X}{}_{CC}$ metrics are products of basic metrics in the same way as for $X_s$ and $X_{CC}$ (with also all compound $A^{X}{}_{Ref} = 1$ by construction), they can similarly be cast if needed into an additive form of logarithmic gains ("$A^{XdB}$ in units [dB]") that sum up to total amplification gain of the compound metric. As noted for the $X_s$, $X_{CC}$ and $X_{Ref}$ metrics above, this can facilitate proper visualization independent of order of multiplication (such as done in Fig. 3a-c and Extended Data Fig. 5a-c for "stacked" amplification factor illustrations for heat extremes). It generally can aid to better deal with strongly amplified extremes, since log-space analysis is often preferable for analyzing composite amplifier gains spanning a factor of 10-100 or more, such as common in fields like acoustics[70] and electronics engineering[71].

In principle also any annual metric $X_p$, or even individual-event metrics like the severity of a particular event of interest (rather than the average event severity $\bar{S}_p$ in an annual CTP), can be normalized to its corresponding $X_{Ref}$ level, to put its degree of amplification into the context of the longer-term changes as quantified by the related $X_s$ and $A^{X}{}_{s}$ metrics and associated individual-year spread and natural variability estimates. This can be of interest, for example, in extreme events attribution[49–51], to help quantify the share of a specific hazard's extremity that is attributable to anthropogenic climate change. In this study we focused on the use of the decadal-mean $A^{X}{}_{s}$ and $A^{X}{}_{CC}$ metrics, however, and employed these as the primary indicators throughout for tracking the extremes of interest (e.g., see Fig. 1b introducing the fundamental amplification-factors chart, and Figs 2-4 all focusing on amplification results).

Given their strict anchoring to unity during the chosen reference period, facilitating transparent quantification of deviations along with straightforward interpretation, we consider the $A^{X}{}_{s}$ metrics also the TEA subclass best suited for assessing climate change amplification vs. "reference climate" and hence use this subclass for co-estimating natural variability (NV). In this study we computed observation-based upper and lower standard spread estimates ($s_{AX}^{NVupp}$, $s_{AX}^{NVlow}$) for the natural variability of the primary metrics $A^{X}{}_{s} \in \{A^{F}{}_{s}, A^{D}{}_{s}, A^{M}{}_{s}, A^{A}{}_{s}; A^{t}{}_{s}, A^{S}{}_{s}, A^{T}{}_{s}\}$ for the example GR Austria, based on long-term historic daily $T^{Max}$ and $P^{24H}$ station data from overall seven representative locations and underpinned by further backward assessment using multicentury historic reanalysis data (see Georegions and datasets, and Extended Data Fig. 2 for example result illustrations).

For a detailed description of the NV estimation see Supplementary Note, section 6.3 therein. Beyond building on long-term historic observations a most common estimation approach is to build on unforced (NV-only) control runs of climate models; in all cases proven methods[49,51] should be followed, especially if serving climate change attribution for operational uses[52,55].

**Computation of decadal-mean aggregate georegion variables.**

Given an AGR definition as a large-scale region comprising a set of eligible cells $l(i,j)$ of a grid of GRs (Europe-scale AGRs in this study; see Georegions and datasets), we build on the available decadal-mean metrics for the eligible GRs of the AGR, $X^{GR} \in \{X_s^{GR}, X_{Ref}^{GR}, X_{CC}^{GR}\}$, and first compute a consistent set of the corresponding AGR-mean metrics $X^{\mathrm{AGR}} \in \{X_s^{\mathrm{AGR}}, X_{\mathrm{Ref}}^{\mathrm{AGR}}, X_{\mathrm{CC}}^{\mathrm{AGR}}\}$ based on cell-area-weighted spatial averaging (weights $w_l = a_l / A^{\mathrm{AGR}}$) and the first-basic-then-compound-metrics approach already used in the previous steps. Subsequently, we compute the associated AGR-mean amplification factor metrics $A^{\mathrm{X,AGR}} \in \{A_s^{\mathrm{X,AGR}}, A_{\mathrm{Ref}}^{\mathrm{X,AGR}}, A_{\mathrm{CC}}^{\mathrm{X,AGR}}\}$ by normalizing the $X^{\mathrm{AGR}}$ metrics to the reference-period mean $X_{\mathrm{Ref}}^{\mathrm{AGR}}$ (i.e., same definition as within the GRs), where again $A_{\mathrm{Ref}}^{\mathrm{X,AGR}} = 1$ by construction for all metrics.

In addition, also involving proper cell-area weighting, we compute upper and lower standard spread estimates of the individual GR values about their AGR mean for all $X^{\mathrm{AGR}}$ and $A^{\mathrm{X,AGR}}$ metrics ($s_{\mathrm{X}}^{\mathrm{AGRupp}}$, $s_{\mathrm{X}}^{\mathrm{AGRlow}}$; $s_{AX}^{\mathrm{AGRupp}}$, $s_{AX}^{\mathrm{AGRlow}}$), associated standard spread estimates for the mean, and 5th and 95th percentile estimates for enabling specification of the 90 % coverage interval for the rather skewed distribution of the individual GR deviates about the AGR mean. For the detailed algorithms for the AGR variables see Supplementary Note, section 7 therein.

In this study the European land AGRs EUR, C-EUR, S-EUR, and N-EUR (Fig. 1c) have been used to obtain overall continent-scale insight into the regional amplification of heat extremes under climate change; see Fig. 4c for illustration of the TEX amplification metrics $A_s^{\mathrm{T,AGR}}$ and $A_{\mathrm{CC}}^{\mathrm{T,AGR}}$ along with 90% CI spread estimates for these AGRs. We also used these AGR results on TEX amplification to intercompare the equivalent AEHC metrics $H_{\mathrm{AEHC,CC}}^{\mathrm{AGR}} = \tilde{C}_{ABL} \cdot \mathcal{T}_{\mathrm{CC}}^{\mathrm{AGR}}$ and in particular $A_{\mathrm{CC}}^{\mathrm{HC,AGR}} = A_{\mathrm{CC}}^{\mathrm{T,AGR}}$ ("AEHC CC vs. Ref") to the amplification of the global-scale northern mid-latitude atmospheric heat content (AHC) increase due to global warming ("AHC gain CC vs. Ref"); see the main text section Amplification of heat extremes and atmospheric heat over Europe and Fig. 4d.

Beyond this study, a typical further AGR definition are country borders, usually obtained from geographic information system (GIS) shapefiles, enabling to similarly derive country-specific insights into TEA metrics of interest, from building on the eligible GR grid cells $l(i,j)$ and associated individual GR results within the selected countries (for example the countries of Europe based on the ERA5 grid of GRs used here). Furthermore, also natural variability estimates obtained at individual GR level (see previous computation step) can be similarly transferred to AGRs, for more aggregate-scale insights if needed for an application.

**Software**

All TEA computations as described are performed with self-developed Python software in a GitLab project TEAmetrics, to be available at https://wegcgitlab.uni-graz.at/arsclisys/teametrics (at the time of this Preprint publication and during review of a subsequent manuscript version, the current TEAmetrics code can be made accessible to interested users on qualified request). The code builds on Python 3, with NumPy, Xarray and Pandas as the main libraries.

The initial public version (access on request) supports multi-scale TEA studies on weather and climate extremes of the type performed in this study (time scales: daily/hourly, annual, decadal-mean & CC-vs-Ref; spatial scales: local-scale input data grid within GRs, regional-scale GRs, aggregate-scale AGRs). It can use scalar key variable data over Europe or other land regions worldwide, with ERA5[31], ERA5-Land[30] and ERA5-HEAT[17] as the primary source datasets to derive key variable fields of interest (e.g., also for impact-oriented variables[3,12,17–19] beyond more conventional ones[12–15]) and associated input data (e.g., reference threshold maps).

Auxiliary data and software tools are all open source as well, such as data access software (e.g., NetCDF library) and geographic information system (GIS) data (e.g., country border shapefiles, land-sea masks, coordinate system conversions between UTM[58] and geodetic/geographic coordinates).

## Data availability

The SPARTACUS v1.5 daily gridded data used in this study cover the period from 1961 to 2023 at a resolution of 1 km × 1 km and are available through the GeoSphere Austria Data Hub at https://data.hub.geosphere.at/dataset/spartacus-v1-1d-1km. The ERA5 | ERA5-Land reanalysis hourly gridded datasets cover the period from 1940 | 1950 to present at a resolution of about 30 km (0.25° × 0.25°) | 10 km (0.1° × 0.1°) and are accessible through the Copernicus Climate Data Store https://cds.climate.copernicus.eu at https://doi.org/10.24381/cds.adbb2d47 (ERA5) and https://doi.org/10.24381/cds.e2161bac (ERA5-Land); the complementary E-OBS v29.0e daily gridded data used for crosscheck of ERA5 results cover the period from 1950 to present at a resolution of 0.25° × 0.25° (and alternatively 0.1° × 0.1°) and can be accessed through the same data store at https://doi.org/10.24381/cds.151d3ec6. The long-term Measurement stations Daily data v2 datasets (stations Graz, Vienna, Kremsmünster, Salzburg, Innsbruck, Bad Gleichenberg, Deutschlandsberg), used in the example GR Austria for natural variability estimates, and the related complementary long-term HISTALP station data, can be accessed through the GeoSphere Austria Data Hub https://data.hub.geosphere.at at https://doi.org/10.60669/gs6w-jd70 (Daily data v2) and https://www.zamg.ac.at/histalp (HISTALP), respectively. The multicentury ModE-RA historic monthly reanalysis data over 1422 to 2008 are accessible via the ClimeApp portal https://mode-ra.giub.unibe.ch/climeapp. The main TEA result metrics for heat and precipitation extremes from the 1960s to the present are available through the Graz Climate Change Indicators portal https://gcci.uni-graz.at/ewm (Extreme Weather Monitoring), also accessible via https://climatetracer.earth (access under preparation at the time of this Preprint publication; becoming available in the coming months during work on a subsequent manuscript version).

## Code availability

The code used to obtain the TEA results of this study is set to become available through a GitLab project TEAmetrics, to be accessible at https://wegcgitlab.uni-graz.at/arsclisys/teametrics (at the time of this Preprint publication and during review of a subsequent manuscript version, the current TEAmetrics code can be made accessible to interested users on qualified request).

## Additional Methods References

58. Iliffe, J. Transverse Mercator projection–Universal Transverse Mercator (UTM) system. in *Datums and Map Projections* (Iliffe, J.) 74–77 (Whittles Publishing, 2005).
59. Cornes, R. C., van der Schrier, G., van den Besselaar, E. J. M. & Jones, P. D. An ensemble version of the E-OBS temperature and precipitation datasets. *J. Geophys. Res. Atmos.* **123**, 9391–9409 (2018).
60. Warren, R. et al. ClimeApp: Opening doors to the past global climate. New data processing tool for the ModE-RA climate reanalysis. Preprint at EGUsphere https://doi.org/10.5194/egusphere-2024-743 (2024).
61. Zhang, X., Zwiers, F. W., Li, G., Wan, H. & Cannon, A. J. Complexity in estimating past and future extreme short-duration rainfall. *Nat. Geosci.* **10**, 255–259 (2017).
62. Fowler, H. J. et al. Anthropogenic intensification of short-duration rainfall extremes. *Nat. Rev. Earth Environ.* **2**, 107–122 (2021).
63. Armstrong McKay, D. I. et al. Exceeding 1.5°C global warming could trigger multiple climate tipping points. *Science* **377**, eabn7950 (2022).
64. Stocker, T. F. et al. Reflecting on the science of climate tipping points to inform and assist policy making and address the risks they pose to society. *Surv. Geophys.* Publ. online https://doi.org/10.1007/s10712-024-09844-w (2024).

65. Schauberger, B. et al. Consistent negative response of US crops to high temperatures in observations and crop models. *Nat. Commun.* **8**, 13931 (2017).
66. Tesfaye, K. et al. Climate change impacts and potential benefits of heat-tolerant maize in South Asia. *Theor. Appl. Climatol.* **130**, 959–970 (2017).
67. Seidel, D. J. et al. Climatology of the planetary boundary layer over the continental United States and Europe. *J. Geophys. Res. Atmos.* **117**, D17106 (2012).
68. Collaud Coen, M. et al. Determination and climatology of the planetary boundary layer height above the Swiss plateau by in situ and remote sensing measurements as well as by the COSMO-2 model. *Atmos. Chem. Phys.* **14**, 13205–13221 (2014).
69. Bakas, N. A., Fotiadi, A. & Kariofillidi, S. Climatology of the boundary layer height and of the wind field over Greece. *Atmosphere* **11**, 910 (2020).
70. Brown, P. Fundamentals of audio and acoustics. in *Handbook for Sound Engineers–Fourth Edition* (Ed Ballou, G.) 21–39 (Elsevier Focal Press, 2008).
71. Hulick, T. P. Solid-state amplifiers. in *The Electronics Handbook–Second Edition* (Ed Whitaker, J. C.) 577–611 (Taylor & Francis, 2005).

## Acknowledgements

We thank M. Gorfer for support in the AHC computations, C. Brimicombe for valuable discussions on extreme heat indices, and M. Pichler for support in linking the TEA metrics to the Graz Climate Change Indicators portal. This work was supported by WegenerNet funding that is provided by the University of Graz based on field of excellence and core research infrastructure funds received from the Austrian Federal Ministry for Education, Science and Research, and by the State of Styria and the City of Graz; detailed information can be found online (https://wegcenter.uni-graz.at/wegenernet). The research is part of the Field of Excellence Climate Change Graz (https://climate-change.uni-graz.at).

## Author contributions

G.K. conceived the TEA methodology and the study approach and design, supported by S.J.H. and J.F. in its refinement and consolidation. S.J.H. performed the original TEA code implementation and the data analysis and figures visualization, guided in this work by G.K. and co-advised by J.F. G.K. wrote the original draft, which was reviewed and edited by S.J.H. and J.F. J.F. implemented the TEAmetrics software, supported by G.K. and S.J.H. Supervision and funding acquisition was by G.K.

## Supplementary Information – Supplementary Note

Kirchengast, G., Haas, S. J. & Fuchsberger, J. Compound event metrics detect and explain ten-fold increase of extreme heat over Europe—Supplementary Note: Detailed methods description for computing threshold-exceedance-amount (TEA) indicators. *Supplementary Information (SI) to Preprint – April 2025.* 40 pp. Wegener Center, University of Graz, Graz, Austria, 2025. Available from the authors on qualified request (e-mail: gottfried.kirchengast@uni-graz.at).

– end of document –